\documentclass[%
 reprint,
preprintnumbers,
nofootinbib,
 amsmath,amssymb,
 aps,
]{revtex4-2}

\usepackage{graphicx}
\usepackage{dcolumn}
\usepackage{bm}

\usepackage{soul}

\usepackage{myPreamble}

\begin{document}

\preprint{TUM-HEP-1540/24}

\title{False vacuum decay beyond the quadratic approximation: summation of non-local self-energies}

\author{Matthias Carosi}
 \email{matthias.carosi@tum.de}
\author{Björn Garbrecht}%
 \email{garbrecht@tum.de}
\affiliation{%
    Physik-Department, Technische Universit\"at M\"unchen,\\
    James-Franck-Str., 85748 Garching, Germany
}%

\date{\today}

\begin{abstract}
Using the 2PI effective action formalism, we study false vacuum decay beyond the quadratic approximation of the path integral. We derive a coupled system of equations for the bounce and the propagator, and we compute a semi-analytic expression for the self-energy of a real scalar field with cubic and quartic interactions from the 2PI effective action truncated at two loops and without further approximations. Deriving numerical results, we can show that the Hartree approximation, where non-local contributions to the self-energy are neglected, is generally not justified. 
The procedure we develop is a key step towards the explicit computation of the quantum corrected bounce, the determinant of fluctuations about it and the decay rate in the presence of classical zero-modes that are lifted by quantum effects, e.g. classically scale-invariant models relevant for assessing the Higgs stability.
\end{abstract}

\maketitle


\section{Introduction}
Tunnelling is a non-perturbative effect unique to quantum theory and arises in the presence of a metastable state in the theory.
A well-established phenomenon in quantum mechanics, tunnelling in quantum field theory (QFT) still awaits experimental observation despite some early positive results for thermally activated transitions~\cite{zenesini2023observation}.

In QFT, a technique for computing the tunnelling rate---or the decay width---of a metastable vacuum was first introduced in the seminal papers by Callan and Coleman~\cite{PhysRevD.15.2929,Callan:1977pt}.
Said method based on instantons, despite some proposed alternatives (e.g. \cite{Espinosa:2019hbm}), remains to this day the most widely used and reliable way to compute such rates. While we partly recapitulate this approach in Section~\ref{sec: decay rate review}, we will omit many finer details on this well-established method and refer the interested reader to some other references~\cite{10.1093/acprof:oso/9780198509233.001.0001,Andreassen_2017}.

Of both theoretical and phenomenological interest~\cite{CABIBBO1979295,Isidori_2001,Andreassen:2017rzq} is the evaluation of the decay rate of the false vacuum of the classically scale invariant scalar theory with potential unbounded from below, whose Euclidean Lagrangian in four dimensions is 
\begin{equation}\label{eqn: euclidean lagrangian of phi4}
    \mathcal{L}_\mathrm{E} = \frac{1}{2} \partial_\mu \varphi \partial_\mu \varphi + \frac{\lambda}{4!}\varphi^4 \,, \qquad \mathrm{where} \qquad \lambda<0\,.
\end{equation}
This model exhibits a metastable vacuum at $\varphi\equiv 0$ that can decay via quantum fluctuations towards the infinite well, described by the Fubini-Lipatov instanton~\cite{Fubini:1976jm,Lipatov:1976ny}.
It presents challenging technical difficulties due to the additional classical symmetry---scale invariance.
Because of this, instantons of all sizes contribute equally to the partition function, leading to a problematic infrared divergence.
Since the classical scale invariance, however, is not a symmetry of the quantum theory, the infrared problem should be solved when accounting for quantum corrections.
Inspired by this fact, renormalisation group (RG) arguments were used in Ref.~\cite{Isidori_2001} to obtain for the first time the decay rate to next-to-leading (NLO) order.

In Ref.~\cite{Andreassen:2017rzq}, the authors refine this argument with the inclusion of the two-loop RG running and the resummation of leading logarithms to all loops, plus the resolution of a subtle, yet crucial issue with the collective coordinate method, namely the appearance of an infinite Jacobian for the dilatational mode.

However, their procedure for regularising the infrared divergence relies critically on the $\beta$-function for the self-coupling~$\lambda$ vanishing at a critical scale~$\mu^*$, such that~${\beta_\lambda(\mu^*) = 0}$. While this is a justified assumption in the context of Higgs vacuum decay in the Standard Model, where the competing effects from the top quark and the weak gauge bosons give rise to said critical scale, it is not true for a generic field theory described by the Lagrangian~\eqref{eqn: euclidean lagrangian of phi4}, possibly plus interaction with additional fields.
Describing the decay of the metastable vacuum at $\varphi\equiv0$ is, however, a well-posed question even in theories where the running of the coupling does not exhibit a stationary point.

Furthermore, while the RG argument can fix the renormalisation scale dependence of the quantum corrections, it is unable to catch their full contribution, which includes deformations perpendicular to dilatations. 
These need to be treated accurately in view of the fact that soft modes remain present for low angular momenta. Especially for the quantum corrected bounce and angular momentum one, the translational zero-modes must be subtracted, which can only be done when going beyond the leading loop approximation.

Another way to formulate the problem is that the quantum bounce (instanton) describing the decay of the vacuum is non-perturbatively far away from the tree-level one.
The critical radius of the bounce solution is only generated once we include quantum corrections, and any approach trying to assess the stability of the vacuum needs a systematic inclusion of quantum effects beyond expansions in small fluctuations about the tree-level bounce.

In order to capture the effect on the fluctuations about the instanton, which is of crucial importance in the presence of a classical symmetry broken by quantum effects, we must sum the leading quantum corrections into the propagator~\cite{Garbrecht:2018rqx}.
This is a key step towards the explicit calculation of the quantum corrections to the decay rate which should remove the infrared problem mentioned above.
To do so, we choose to work in the framework of the two-particle irreducible (2PI) effective action, which provides a natural setting to generate a consistent system of equations for the one- and two-point functions including loop effects.
Starting from the two-loop 2PI effective action, we derive the quantum equation of motion for the instanton and the one-loop Schwinger-Dyson equation for the Green's function in the instanton background.
This results in a system of integro-differential equations, which we solve self-consistently for a simplified two-dimensional model for the first time without further approximations, such as discarding non-local corrections in the form of convolution integrals.
Once the solution of the equations is known, the decay rate can be fully expressed in terms of the instanton and the Green's function, though we leave the calculation of it to future work.

The paper is structured as follows. 
In Section~\ref{sec: decay rate review}, we formulate the problem of finding the decay rate purely in terms of the effective action.
In Section~\ref{sec: fv decay in 2pi}, we use the 2PI effective action formalism and obtain a system of coupled non-linear integro-differential equations for the one and two-point functions, namely for the background and the propagator.
For the first time, in Section~\ref{sec: self-energy} we find a closed-form expression for the full one-loop self-energy of a real scalar field with cubic and quartic interaction in a radially symmetric background, including non-local diagrams. We discuss at length the subtleties related to divergences and present a fully renormalised result in four dimensions.
All of our results up to this point are fully general and applicable to any real scalar theory such as the model in Eq.~\eqref{eqn: euclidean lagrangian of phi4}.
To give a working proof of the procedure we outline, in Section~\ref{sec: parametric example d=2}, we present numerical results for a two-dimensional example. We show that the non-local contribution to the self-energy can, in general, not be ignored, namely that the Hartree approximation is generally not justified.
We find that while quantum corrections leave the one-point function almost unchanged, they affect the propagator and the fluctuation eigenvalues. This leads us to expect similar effects for the fluctuation determinant appearing in the evaluation of the decay rate.
We should stress that we focus on a simplified model which does not exhibit classical scale invariance and, thus, does not suffer from the infrared problem we are aiming to address eventually.
The numerical application to a more realistic case study is left for a future endeavour.
Finally, in Section~\ref{sec: conclusions}, we make some concluding remarks.

\section{The decay rate from the effective action: a brief review}\label{sec: decay rate review}
Consider the theory of a single real scalar field~$\varphi$ in~$d$-dimensional Euclidean space, described by the action
\begin{equation}
    S[\varphi] = \: \int \d^d x \, \left( \frac{1}{2} \partial_\mu \varphi \partial_\mu \varphi + U(\varphi) \right) \,, \label{eqn: action generic scalar theory}
\end{equation}
where the index~$\mu$ goes from~1 to~$d$.
The potential~$U(\varphi)$ is characterised by having one local minimum~$\varphi_{\mathrm{FV}}$ which we call the false vacuum, and a global minimum~$\varphi_{\mathrm{TV}}$ called the true vacuum. Both of them satisfy~${U'(\varphi_{\mathrm{FV}}) = U'(\varphi_{\mathrm{TV}}) = 0}$, and~${U(\varphi_{\mathrm{FV}}) > U(\varphi_{\mathrm{TV}})}$.

We denote by $V$ the spatial volume of the system under consideration and by $T$ the Euclidean time. 
By the optical theorem, the decay rate per unit volume of a homogeneous field configuration in the false vacuum~${\varphi\equiv\varphi_{\mathrm{FV}}}$ is given by
\begin{equation}
    \frac{\mathit{\Gamma}_{\mathrm{FV}}}{V} = \: - \frac{2 \im E_{\mathrm{FV}}}{V} = \: \lim_{T \to \infty} \frac{2}{V T} \im \log Z_E\,,
\end{equation}
where we have used that in the infinite Euclidean time (zero temperature) limit, everything is projected onto the ground state.
The Euclidean partition function~$Z_E$ can be written as a path integral
\begin{equation}\label{eqn: euclidean partition function PI}
    Z_E = \: \mathcal{N} \hspace{-1em} \underset{\varphi(r\to\infty) = \varphi_{\mathrm{FV}}}{\int} \hspace{-1em} \left[\mathcal{D}\varphi \right] \, e^{-S[\varphi]} \,,
\end{equation}
where~${r^2 = x_\mu x_\mu}$ is the square of the radius in $d$-dimensional spherical coordinates. 
The boundary conditions ensure that we only consider configurations that sit in the false vacuum asymptotically. This follows the intuition by Coleman and then by Callan and Coleman~\cite{PhysRevD.15.2929,Callan:1977pt}, that the Euclidean path integral with such boundary conditions receives an imaginary contribution due to a non-trivial saddle point of the action, called \textit{the bounce}. This is an $O(d)$-symmetric field configuration, found by solving the equation of motion
\begin{equation}
    \frac{\delta S}{\delta \varphi_b} = \: - \frac{1}{r^{d-1}} \frac{\d}{\d r} r^{d-1} \frac{\d}{\d r} \varphi_b(r) + U'(\varphi_b(r)) = \: 0 \,,
\end{equation}
together with the requirement~${\lim_{r\to0} \d \varphi_b /\d r = 0}$ ensuring that the solution is non-singular, plus the boundary condition of the path integral~${\lim_{r\to\infty} \varphi_b(r) = \varphi_{\mathrm{FV}}}$.

Recently~\cite{Andreassen_2017,Andreassen:2017rzq}, the existence of an additional saddle point has put into question the validity, or at least the derivation, of the instanton method. 
This discussion, however, goes beyond the purpose of this work, and we assume that a solution to this issue exists~\cite{Andreassen_2016}.

Generally, one can evaluate the path integral in the partition function following steepest descent contours emerging from saddle points. This way, the Laplace approximation can be used, which is tantamount to the semiclassical expansion around the saddle points. To address vacuum decay, we must include the contour ${\cal J}_{\rm FV}$ pertaining to the false vacuum as well as ${\cal J}_b$ pertaining to the bounce. Note that the contour ${\cal J}_{\rm FV}$ is not unbounded but ends in the bounce, where a Stokes phenomenon is encountered. We also must account for multi-bounces as will be discussed further down the line.

Expanding the Euclidean partition function from Eq.~\eqref{eqn: euclidean partition function PI} in the saddle point approximation
\begin{equation}
    Z_E = \: \mathcal{N} \int_{\mathcal{J}_{\mathrm{FV}}} [\mathcal{D}\eta] \, e^{-S[\varphi_{\mathrm{FV}} + \eta]} + \frac{1}{2} \mathcal{N}\int_{\mathcal{J}_{b}} [\mathcal{D}\eta] \, e^{-S[\varphi_b + \eta]} + \ldots\,,
\end{equation}
where we have denoted with~$\varphi_b(x)$ the bounce, and with~${\eta(x)}$ the fluctuations around either background. The omitted terms in this expansion are the contributions coming from multi-bounces, which we will re-introduce shortly.
The factor of~$1/2$ comes from the fact that we are only integrating over half of the steepest descent contour associated with the bounce, as a consequence of the aforementioned Stokes phenomenon. 

Now, we recall that the effective action can be defined via the background field identity (see e.g.~\cite{Berges_2002}) \begin{equation}
    e^{-\Gamma_{\mathrm{eff}}[\varphi]} = \: \int_{\mathcal{J}_{\overline{\varphi}}} [\mathcal{D} \eta ] e^{-S[\varphi + \eta] + \int_x \frac{\delta \Gamma_{\mathrm{eff}}[\varphi]}{\delta \varphi_x} \eta_x} \,,
\end{equation}
which, when computed on a saddle point $\overline\varphi$ of the effective action, reduces to 
\begin{equation}
    e^{-\Gamma_{\mathrm{eff}}[\overline{\varphi}]} = \: \int_{\mathcal{J}_{\overline{\varphi}}} [\mathcal{D} \eta ] e^{-S[\overline{\varphi} + \eta]} \,.
\end{equation}
Note that $\Gamma_{\mathrm{eff}}[\overline{\varphi}]$ is not the full effective action but only the one associated with the steepest descent contour~${\cal J}_{\overline\varphi}$. The full effective action is obtained when summing the according contributions from all contributing saddle points.

For non-homogeneous field configurations, the story is slightly more complicated as we must consider the presence of zero modes due to the breaking of symmetries.
For the bounce, for example, we must remember that we do not just have one single saddle-point, but rather a~$d$-dimensional manifold corresponding to the~$d$-parameter family of solutions $\varphi_b^{x_0}$. The moduli~${x_0 = (x_0^1, \ldots, x_0^d)}$ parametrise the location of the centre of the bounce in the $d$-dimensional space. The configuration~$\varphi_b^{x_0}$ is then a valid saddle point for any~$x_0$, which implies the existence of a zero mode among the fluctuations~$\eta$, corresponding to just changing the value of~$x_0$. This is the manifestation of translational symmetry, which the theory possesses but the bounce breaks.
We must integrate out the zero modes, and we do so by the method of collective coordinates
\begin{align}
    \int [\mathcal{D}\eta] e^{-S[\varphi_b+\eta]} & = \: \int_{\mathbb{R}^d} \d x_0 J_{\mathrm{tr}} \int [\mathcal{D}\eta]' e^{-S[\varphi_b+\eta]} \notag\\[2ex]
    & = \: VT \left(\frac{||\partial_\mu\varphi_b||^2}{2\pi}\right)^{\frac{d}{2}} e^{-\Gamma_{\mathrm{eff}} [\varphi_b]}\,. \label{eqn: collective coordinates for translations}
\end{align}
The prime indicates that the translational zero modes are dropped from the integration and~${J_{\rm tr}=(||\partial_\mu\varphi_b||^2/2\pi)^{d/2}}$ is the Jacobian that comes from trading these modes in favour of the ${\rm d}x_0$-integral. We shall also use the notation with the prime to indicate the omission of zero modes in determinants.

We can rewrite the partition function in the saddle point approximation as
\begin{align}
    Z_E =& \: \mathcal{N} e^{-\Gamma_{\mathrm{eff}}[\varphi_{\mathrm{FV}}]}\\\notag\times& \left( 1 + \frac{VT}{2}J_{\mathrm{tr}} e^{-\Gamma_{\mathrm{eff}}[\varphi_b] + \Gamma_{\mathrm{eff}}[\varphi_{\mathrm{FV}}]} + \ldots \right) \,.
\end{align}
Now, it is time to include the contribution of multi-bounces. 
The saddle point approximation relies on the coupling being weak, which in turn justifies the dilute instanton gas approximation.
Then we can view a multi-bounce configuration as a collection of widely separated individual bounces freely located in spacetime.
Then, for each bounce, we have a factor of~$VT J_{\rm tr}$ coming from the zero modes, and a factor of~$1/2$ for the negative mode.
The effective action of a multi-bounce configuration in the dilute approximation is just given by the sum of the effective action of each individual bounce, where we must take care that we do not over-count the contribution of the false vacuum by including it only once.
The contribution of the~$n$ bounce configuration to the partition function then reads
\begin{equation}
    \mathcal{N} \frac{1}{n!} \left(\frac{VT}{2} J_{\mathrm{tr}} \right)^n e^{-n(\Gamma_{\mathrm{eff}}[\varphi_b] - \Gamma_{\mathrm{eff}}[\varphi_{\mathrm{FV}}]) - \Gamma_{\mathrm{eff}}[\varphi_{\mathrm{FV}}]}\,,
\end{equation}
and summing over all contributions we obtain
\begin{equation}
    \log Z_E = -\Gamma_{\mathrm{eff}} [\varphi_{\mathrm{FV}}] + \frac{VT}{2} J_{\mathrm{tr}} e^{-\Gamma_{\mathrm{eff}}[\varphi_b] + \Gamma_{\mathrm{eff}}[\varphi_{\mathrm{FV}}]}\,.
\end{equation}
When taking the imaginary part only the contribution of the bounce survives and we find 
\begin{equation}\label{eqn: decay rate eff action}
    \frac{\mathit{\Gamma}_{\mathrm{FV}}}{V} = \: J_{\rm tr} \left|e^{-\Gamma_{\mathrm{eff}}[\varphi_b] + \Gamma_{\mathrm{eff}}[\varphi_{\mathrm{FV}}]} \right|\,.
\end{equation}
Thus, we have expressed the decay rate of a metastable vacuum only in terms of the effective action computed around the quantum bounce, obtained as saddle point of the full quantum effective action.

To relate to the more familiar expression, let us take $\Gamma_{\mathrm{eff}}[\varphi]$ to be the one-particle irreducible (1PI) effective action and expand it to one loop
\begin{equation}
    \Gamma_{1\mathrm{PI}} [\varphi] = \: S[\varphi] + \frac{\hbar}{2} \tr \log G_0^{-1}(\varphi) + \mathcal{O}(\hbar^2) \,,
\end{equation}
where $G_{0,\varphi}$ is the tree-level propagator and its inverse is the quadratic part of the classical action computed around the background field~$\varphi$
\begin{equation}\label{eqn: inverse quadratic action}
    G_{0,\varphi}^{-1} (x,y) = \frac{\delta^2 S[\varphi + \eta]}{\delta\eta(x) \delta\eta(y)}\Big\rvert_{\eta \equiv 0}\,.
\end{equation}
Plugging this into Eq.~\eqref{eqn: decay rate eff action} we obtain the well-known result first found by Callan and Coleman~\cite{Callan:1977pt}
\begin{align}
    \frac{\mathit\Gamma_{\rm FV}}{V} = \: & \left(\frac{S[\varphi_b]}{2\pi \hbar}\right)^{\frac{d}{2}} \left| \frac{\sideset{}{'}\det G_0^{-1}(\varphi_b)}{\det G_0^{-1}(\varphi_{\mathrm{FV}}) } \right|^{-\frac{1}{2}} \times \notag \\[1.5ex]
    & \qquad \times e^{-\frac{1}{\hbar} (S[\varphi_b]-S[\varphi_{\mathrm{FV}}]) } \left(1 + \mathcal{O}(\hbar)\right) \,,
\end{align}
where we have re-inserted $\hbar$ to make the counting in loop order more explicit, and we used~${||\partial_\mu\varphi_b||^2=S[\varphi_b]+\mathcal{O}(\hbar)}$ in the Jacobian of translations.
Note that the translational zero modes have already been subtracted from the theory in Eq.~\eqref{eqn: collective coordinates for translations}. This means that when inverting the quadratic part of the action from Eq.~\eqref{eqn: inverse quadratic action} in the background of the bounce to obtain the tree-level propagator~$G_{0,\varphi_b}$, we should only do so on the subspace of the Hilbert space complementary to the span of the translational zero modes. 
In equations, this translates into the following Green's function equation
\begin{align}
     \int \d^d y \: \frac{\delta^2 S[\varphi + \eta]}{\delta\eta(x) \delta\eta(y)}\Big\rvert_{\eta \equiv 0} \: G_{0,\varphi_b} (y,z) = \: & \delta^{(d)}(x-z) \notag \\
     & \hspace{-3em} - \phi_\mu(x) \phi_\mu(z) \,, \label{eqn: definition subtracted green's function}
\end{align}
where~$\phi_\mu$ for~${\mu=1,\ldots,d}$ are the~$d$ translational zero modes.
The right-hand side of Eq.~\eqref{eqn: definition subtracted green's function} is the identity on the subspace orthogonal to the zero modes.

To go beyond the quadratic approximation, we can expand the 1PI effective action to higher loop orders and solve for the quantum bounce solution to the quantum equation of motion, namely
\begin{align}
    -\frac{1}{r^{d-1}} \frac{\d}{\d r} r^{d-1} \frac{\d}{\d r} \varphi_b(r) + U'(\varphi_b(r)) & \notag \\
    + \int_{r'} \mathbf{\Pi}_b (r,r') \varphi_b(r') & = \: 0\,, \label{eqn: 1pi quantum eom}
\end{align}
where $\mathbf{\Pi}_b$ is the self-energy computed in the $\varphi_b$ background that solves the above equation.
In particular,~$\mathbf{\Pi}_b$ is the sum of 1PI diagrams from tree-level propagators~$G_{0,\varphi_b}$ computed in the background of the field configuration~$\varphi_b$ by solving Eq.~\eqref{eqn: definition subtracted green's function}.
There are two problems with this approach.

The first one is that what is the zero mode in the classical theory, namely the radial derivative of the bounce, cannot be a zero mode of the operator~(\ref{eqn: inverse quadratic action}) any more once~$\varphi_b$ is the quantum bounce. In fact, by differentiating the quantum equation of motion~\eqref{eqn: 1pi quantum eom}, we clearly do not recover the tree-level zero-mode equation.
This is a well-known feature of truncations of the effective action: by truncating at a certain loop level, the conditions for Goldstone's theorem are not met, and the would-be Goldstone mode of translations acquires a non-zero mass, or here, eigenvalue.
To tackle this, various techniques have been developed, in particular the so-called symmetry improved effective action as introduced in Ref.~\cite{Pilaftsis:2013xna} and further studied in Ref.~\cite{Garbrecht:2015cla}.
Note that this implies a modification of \emph{all} eigenvalues, in particular the negative one, which might be shifted to some positive value.

The second issue we face is that quantum corrections to the propagator, and therefore to the eigenvalues, in a non-trivial background, such as the bounce, can be expected to be rather large. This, in turn, implies large corrections to the next-to-leading order term in the effective action, namely the functional determinant over the fluctuation modes.
This effect is especially strong in theories where the quantum bounce is non-perturbatively far away from the tree-level solution, as is the case for the theory described by the Lagrangian~\eqref{eqn: euclidean lagrangian of phi4} where classical scale invariance is broken by quantum corrections.
By only using tree-level propagators, we are thus prone to large errors, at least in some regions of parameter space.
This phenomenon has been first reported in Refs.~\cite{Bergner:2003id,Baacke:2004xk,Baacke:2006kv}, employing the 2PI effective action in the Hartree approximation, namely where all non-local contributions to the self-energy are ignored.

In the present work, we put aside the first issue, for which approaches have been developed and applied in the context of the Hartree approximation for false vacuum decay in Ref.~\cite{Garbrecht:2018rqx}. We focus instead on developing a systematic summation scheme for the quantum corrections to the propagator in the 2PI effective action formalism. 
The application in Ref.~\cite{Garbrecht:2015yza} focuses on large deformations of the potential that appear in the Hartree approximation and for thin-wall bubbles. Here, we in particular go beyond the Hartree approximation and include the non-local contributions in the self-energy without relying on the thin-wall approximation.
As we will demonstrate, the non-local diagrams can be as relevant as the local ones and thus require to be accounted for on the same footing.

Further recent context can be found in Ref.~\cite{Batini:2023zpi}, where the real-time dynamics of false vacuum decay in the 2PI effective action formalism at large~$N$ is investigated. 
The authors study the transition by comparing it to high-temperature statistical field theory simulations on the lattice. In the present work, instead, we focus on a semi-analytical approach to the zero-temperature transition described in terms of Euclidean instantons.


\section{False vacuum decay in the 2PI effective action formalism}\label{sec: fv decay in 2pi}
While we do not re-introduce the formalism of the 2PI-effective action~\cite{Cornwall:1974vz}---for a review see Ref.~\cite{Berges_2004}---we shall recall here a few key steps useful for our arguments, see also Ref.~\cite{Garbrecht:2015yza}.
The 2PI effective action is a functional of the exact~one and two-point functions~$\varphi$ and~$G$, and it can be written in a loop expansion as
\begin{equation}
    \Gamma_{2\rm PI} [\varphi, G] = \: S[\varphi] + \frac{\hbar}{2} \tr G_{0,\varphi}^{-1} G + \frac{\hbar}{2} \tr \log G^{-1} + \Gamma_2[\varphi,G]\,, \label{eqn: 2pi effective action}
\end{equation}
where~$\Gamma_2$ is minus the sum of all 2PI vacuum diagrams starting at two loops. Note that always~$G$, the exact (dressed) propagator, appears in the diagrams.
We work in renormalised perturbation theory, meaning that all of the masses and couplings that we write are to be understood as the renormalised ones, and the counter-terms appear explicitly in the action~$S$, though we will avoid writing them. Furthermore,~$\Gamma_2$ contains diagrams with counter-terms as vertices as well.
For more details on the renormalisation of the 2PI effective action, see Ref.~\cite{Berges_2005}.

From the effective action~\eqref{eqn: 2pi effective action}, we can easily generate equations of motion for the background and for the propagator.
Taking the variation of~$\Gamma_{2\rm PI}$ with respect to the background field, we find the equation of motion for the one-point function
\begin{equation}
    - \laplacian \varphi (x) + U' (\varphi(x)) +\frac{\hbar}{2} U'''(\varphi(x)) G(x,x) + \frac{\delta \Gamma_2[\varphi,G]}{\delta \varphi(x)} = \: 0 \,.\label{eqn: 2pi eom full}
\end{equation}
Similarly, for the two-point function,
\begin{equation}
    G_{0,\varphi}^{-1}(x,y) - G^{-1}(x,y) + \frac{2}{\hbar} \frac{\delta \Gamma_2[\varphi,G]}{\delta G(x,y)} = \: 0 \,, \label{eqn: 2pi inverse green eq}
\end{equation}
which can be recast into
\begin{align}
    \int \d^d y & \left(G_{0,\varphi}^{-1} (x,y) +  \mathbf{\Sigma}_\varphi(x,y) \right) G(y,z) \notag \\
    &\qquad \qquad = \: \delta^{(d)}(x-z) - \phi_\mu(x) \phi_\mu (z) \,, \label{eqn: 2pi eq G full}
\end{align}
where~$\phi_\mu$ are the~$d$ translational zero modes in the bounce background.
We have defined the self-energy
\begin{equation}\label{eqn: definition self-energy from gamma 2}
    \mathbf{\Sigma}_\varphi(x,y) = \: \frac{2}{\hbar} \frac{\delta \Gamma_2[\varphi,G]}{\delta G(x,y)} \,,
\end{equation}
which starts at~$\mathcal{O}(\hbar)$. Note that the solution for $G$ will include the insertions of $\mathbf{\Sigma}_\varphi$. We will therefore occasionally refer to $G$ as the summed or dressed propagator.

Equations~\eqref{eqn: 2pi eom full} and~\eqref{eqn: 2pi eq G full} must be simplified to make any progress.
Recall that we are looking for an~$O(d)$-invariant solution to the equation of motion, namely~${\varphi(x) = \varphi(|x|) = \varphi(r_x)}$.
Then, we have
\begin{equation}
    G_{0,\varphi}^{-1}(x,y) = \: \delta^{(d)} (x-y) \left( - \laplacian_x + U''(\varphi(r_x)) \right)\,.
\end{equation}
Using the spherical symmetry of the problem, we can make the following ansatz for the Green's function:
\begin{equation}
    G(x,y) = \: \frac{1}{(r_x r_y)^\kappa}\sum_{j,\{\ell\}} Y_{j,\{\ell\}}\left( \vec e_x \right) Y_{j,\{\ell\}}\left( \vec e_y \right) G_j(r_x,r_y)\,,
\end{equation}
where we have introduced the shorthand notation~${\kappa = d/2-1}$. The choice for the pre-factor as a power of the radius is arbitrary and is made for later convenience.
We denote by~$\vec e_x$ the $d$-dimensional unit vector parallel to~$x$ and accordingly for~$y$.
Here and in the following, the subscripts~$x$ and~$y$ shall be attached on $r$ to avoid ambiguities whenever there is more than one coordinate in the game.
Recall the sum rule for hyper-spherical harmonics
\begin{equation}
    \sum_{ \{\ell\} } \, Y_{j, \{\ell\}} \left(\vec e_x\right) Y_{j, \{\ell\}} \left(\vec e_y\right) = \: \frac{1}{2\pi^{\kappa+1}} (j+\kappa) \Gamma(\kappa)\, C_j^{\kappa}\left(\cos\theta\right)\,, \label{eqn: sum rule harmonics gegenbauer}
\end{equation}
where~$C_j^\kappa(z)$ are the Gegenbauer polynomials.
Strictly speaking, this sum rule only holds for~$\kappa>0$, namely for~$d>2$. However, only the derivation in Section~\ref{subsec: the bubble} relies on this step. We comment more on this later.
Then the Green's function takes the more amenable form
\begin{equation}\label{eqn: definition Green's fnct in angular decomposition}
    G(x,y) = \: \frac{\Gamma(\kappa)}{2\pi^{\kappa+1}(r_x r_y)^\kappa} \sum_{j=0}^\infty (j+\kappa) \, C_j^{\kappa}\left(\cos\theta\right) \, G_j(r_x,r_y)\,, 
\end{equation}
where we have defined the relative angle between~$x$ and~$y$ via~${\vec e_x \cdot \vec e_y = \cos\theta}$.
As for the self-energy, we split it into a local and a non-local contribution
\begin{equation}
    \mathbf{\Sigma}_\varphi (x,y) = \: \delta^{(d)}(x-y) \Pi_\varphi (r_x) + \Sigma_\varphi(x,y) \,,
\end{equation}
and we again use spherical symmetry to make the ansatz for the non-local part
\begin{align}
    \Sigma_\varphi(x,y) = \: & \frac{1}{(r_x r_y)^\kappa}\sum_{j,\{\ell\}} Y_{j,\{\ell\}}\left(\vec e_x\right) Y_{j,\{\ell\}}\left(\vec e_y\right) \Sigma_{\varphi,j}(r_x,r_y) \notag \\
    = \: & \frac{\Gamma(\kappa)}{2\pi^{\kappa+1}(r_x r_y)^\kappa} \sum_{j=0}^\infty (j+\kappa) \notag \\
    & \qquad \qquad \times C_j^{\kappa}\left(\cos\theta\right) \, \Sigma_{\varphi,j}(r_x,r_y) \,. \label{eqn: ansatz bubble factorisation}
\end{align}

Before re-writing the equations of motion, we make one further simplification: we truncate both equations at~$\mathcal{O}(\hbar)$, meaning $\Gamma_2$ completely drops out from the equation for~$\varphi$, while only~two-loop diagrams from~$\Gamma_2$ will contribute to the self-energy~$\mathbf{\Sigma}_\varphi$.
Note that taking a variation with respect to the propagator~$G$ lowers by one the number of loops of a diagram, meaning we only keep~one-loop diagrams contributing to~$\mathbf{\Sigma}_\varphi$.
However, since~$G$ is the dressed propagator,~$\mathbf{\Sigma}_\varphi$ effectively contains infinitely many loops.
Throwing away~$\Gamma_2$ from Eq.~\eqref{eqn: 2pi eom full}, we find the equation of motion for the background field truncated at~$\mathcal{O}(\hbar)$
\begin{align}
    -\frac{1}{r^{d-1}} \frac{\d}{\d r} r^{d-1} \frac{\d}{\d r} \varphi(r) + U'(\varphi(r)) + \mathfrak{S}_\varphi (r) \varphi(r) = \: 0\,, \label{eqn: 1 loop 2pi eom phi}
\end{align}
We have defined the renormalised~$\mathcal{O}(\hbar)$ term in Eq.~\eqref{eqn: 2pi eom full}
\begin{equation}
    \left[\frac{\hbar}{2} U'''(\varphi(r))G(x,x)\right]_{\mathrm{ren}} = \mathfrak{S}_\varphi(r)\,.
\end{equation}

As for the equation for the propagator, it factorises into a system of equations, one for each angular momentum~$j$
\begin{align}
    &\left( - \frac{1}{r} \frac{\d}{\d r} r \frac{\d}{\d r} + \frac{(j+\kappa)^2}{r^2} + U''(\varphi(r)) + \Pi_\varphi(r) \right) G_j(r,r') \notag \\
    &\quad + \int_0^\infty \d r'' \, r'' \, \Sigma_{\varphi,j}(r,r'') \, G_j(r'',r') \notag \\
    & \qquad = \: \frac{1}{r}\delta(r-r') - \delta_{j,1} r^\kappa \phi_{\mathrm{tr}}(r) \,(r')^\kappa \phi_{\mathrm{tr}} (r') \,. \label{eqn: 1 loop 2pi green eom}
\end{align}
To arrive at Eq.~\eqref{eqn: 1 loop 2pi green eom}, we have used that the hyper-spherical harmonics are a set of orthonormal eigenvectors of the Laplace--Beltrami operator on the~$d-1$ sphere, namely the angular part of the Laplacian in Eq.~\eqref{eqn: 2pi eq G full}.
Also, we rewrote the product of zero modes~${\phi_\mu(x) \propto \partial_\mu \varphi(r)}$ as
\begin{align}
    \sum_{\mu=1}^d \frac{\partial\varphi(r)}{\partial x^\mu} & \frac{\partial\varphi(r')}{\partial x^{\prime \,\mu}} = \: \sum_{\mu=1}^d \frac{x_\mu}{r} \frac{x'_\mu}{r'} \frac{\d \varphi(r)}{\d r} \frac{\d \varphi(r')}{\d r'} \notag \\
    = \: & \sum_{\{\ell\}} Y_{1\{\ell\}}\left( \vec e_x \right) Y_{1\{\ell\}}\left( \vec e_{x'} \right) \phi_{\mathrm{tr}}(r) \phi_{\mathrm{tr}}(r') \,,
\end{align}
which also serves as the definition of the translational zero mode~$\phi_{\mathrm{tr}}$ in the angular momentum basis.

Contrary to what it might look, Eq.~\eqref{eqn: 1 loop 2pi green eom} is a system of \textit{coupled} equations, since both~$\Pi_\varphi$ and~$\Sigma_{\varphi,j}$ depend on all partial waves~$G_j$ simultaneously. This will become more clear in Section~\ref{sec: self-energy}, where we derive explicit expressions for both the local and non-local contributions in a specific theory.

Before that, we highlight that Eqs.~\eqref{eqn: 1 loop 2pi eom phi} and~\eqref{eqn: 1 loop 2pi green eom} are coupled and therefore must be solved simultaneously \textit{and} self-consistently.

\section{Contributions to the self-energy}\label{sec: self-energy}

\subsection{Diagrammatic expansion}

We will focus on the theory described by the action~\eqref{eqn: action generic scalar theory} with a generic cubic and quartic interaction
\begin{equation}\label{eqn: generic potential}
    U(\varphi) = \: \frac{m^2}{2}\varphi^2 - \frac{g}{3!} \varphi^3 + \frac{\lambda}{4!} \varphi^4 \,,
\end{equation}
particularly relevant for~${d=4}$.
This theory has a minimum at~$\varphi=0$, which is a false vacuum if~${g>m\sqrt{18\lambda}}$.
Given potential~\eqref{eqn: generic potential}, we can re-write the~$\mathcal{O}(\hbar)$ term appearing in Eq.~\eqref{eqn: 1 loop 2pi eom phi}
\begin{equation}\label{eqn: 1-loop term in phi eom}
    \mathfrak{S}_\varphi(r) = \: \left[ \frac{\lambda}{2} \varphi(r) G(x,x)\right]_\mathrm{ren} = \: \Pi_\varphi(r) \varphi(r) \,,
\end{equation}
where we have dropped terms in~$U'''(\varphi)$ that are independent of~$\varphi$, as they can be absorbed into the counter-term for the current~$\delta J$.
We call~$\Pi_\varphi$ the \emph{tadpole}, which we soon discuss more in-depth.

There are four diagrams at two-loop order contributing to $\Gamma_2$
\begin{equation} \label{eqn: diagrammatic gamma 2}
    \begin{tikzpicture}
    \draw (0,0) node {$-\Gamma_2[\varphi,G] \supset $};
    \begin{scope}[shift={(2,0)}]
    \draw[very thick] (0,0.5) arc(0:360:0.5);
    \draw[very thick] (0,-0.5) arc(0:360:0.5);
    \end{scope}
    \draw (2.25,0) node {+};
    \begin{scope}[shift={(3,0)}]
    \draw[very thick] (-0.5,0) arc(180:0:0.5);
    \draw[very thick] (0.5,0) arc(0:-180:0.5);
    \draw[very thick] (-0.5,0) -- (0.5,0);
    \end{scope}
    \draw (3.75,0) node {+};
    \begin{scope}[shift={(4.5,0)}]
    \draw[very thick] (-0.5,0) arc(180:0:0.5);
    \draw[very thick] (0.5,0) arc(0:-180:0.5);
    \draw[very thick] (-0.5,0) -- (0.5,0);
    \draw (0.5,0) -- (0.67,-0.12);
    \draw (0.7,-0.2) node[cross=2pt,rotate=0] {};
    \draw (0.7,-0.2) circle (0.1);
    \end{scope}
    \draw (5.5,0) node {+};
    \begin{scope}[shift={(6.5,0)}]
    \draw[very thick] (-0.5,0) arc(180:0:0.5);
    \draw[very thick] (0.5,0) arc(0:-180:0.5);
    \draw[very thick] (-0.5,0) -- (0.5,0);
    \draw (0.5,0) -- (0.67,-0.12);
    \draw (0.7,-0.2) node[cross=2pt,rotate=0] {};
    \draw (0.7,-0.2) circle (0.1);
    \draw (-0.5,0) -- (-0.67,-0.12);
    \draw (-0.7,-0.2) node[cross=2pt,rotate=0] {};
    \draw (-0.7,-0.2) circle (0.1);
    \end{scope}
    \end{tikzpicture}
    \,,
\end{equation}
where each propagator line is the Green's function~$G$, and each cross is a field insertion~$\varphi$.
By taking the variation of $\Gamma_2$ with respect to the propagator~$G$, we get the one-loop self-energy as defined in Eq.~\eqref{eqn: definition self-energy from gamma 2} including both local and non-local contributions
\begin{equation}\label{eqn: diagrammatic self-energy}
    \begin{tikzpicture}
    \draw (0,0) node {$\mathbf\Sigma_\varphi \supset $};
    \begin{scope}[shift={(1.25,0)}]
    \draw[very thick] (0,0.5) circle (0.5);
    \fill (0,0) circle (0.1);
    \end{scope}
    \draw (2,0) node {+};
    \begin{scope}[shift={(3,0)}]
    \fill (-0.5,0) circle (0.1);
    \draw[very thick] (0,0) circle (0.5);
    \fill (0.5,0) circle (0.1);
    \end{scope}
    \draw (4,0) node {+};
    \begin{scope}[shift={(5,0)}]
    \fill (-0.5,0) circle (0.1);
    \draw[very thick] (0,0) circle (0.5);
    \fill (0.5,0) circle (0.1);
    \draw (0.5,0) -- (0.67,-0.12);
    \draw (0.7,-0.2) node[cross=2pt,rotate=0] {};
    \draw (0.7,-0.2) circle (0.1);
    \end{scope}
    \draw (1,-1.5) node {+};
    \begin{scope}[shift={(2.2,-1.5)}]
    \fill (-0.5,0) circle (0.1);
    \draw[very thick] (0,0) circle (0.5);
    \fill (0.5,0) circle (0.1);
    \draw (-0.5,0) -- (-0.67,-0.12);
    \draw (-0.7,-0.2) node[cross=2pt,rotate=0] {};
    \draw (-0.7,-0.2) circle (0.1);
    \end{scope}
    \draw (3.1,-1.5) node {+};
    \begin{scope}[shift={(4.2,-1.5)}]
    \fill (-0.5,0) circle (0.1);
    \draw[very thick] (0,0) circle (0.5);
    \fill (0.5,0) circle (0.1);
    \draw (0.5,0) -- (0.67,-0.12);
    \draw (0.7,-0.2) node[cross=2pt,rotate=0] {};
    \draw (0.7,-0.2) circle (0.1);
    \draw (-0.5,0) -- (-0.67,-0.12);
    \draw (-0.7,-0.2) node[cross=2pt,rotate=0] {};
    \draw (-0.7,-0.2) circle (0.1);
    \end{scope}
    \end{tikzpicture}
    \,,
\end{equation}
where we suppress symmetry factors for simplicity.
We easily identify two types of diagrams, which hereon we call the \emph{tadpole} and the \emph{bubble}, i.e. the local and non-local contribution, respectively. Note from Eq.~\eqref{eqn: 1-loop term in phi eom} that the former also appears in the equation of motion for the background.
In the remainder of this section, we derive explicit expressions for both contributions in terms of the Green's function~$G$.
Then, we discuss the renormalisation of both types of diagram, with a focus on the~${d=4}$ case.

\subsection{The tadpole}\label{subsec: the tadpole}
Up to renormalisation, which we will discuss shortly, the tadpole reads
\begin{align}
    \Pi_\varphi (x) = \: & \frac{\lambda}{2} G(x,x) \notag \\
    = \: & \displaystyle \frac{\lambda}{2} \frac{\Gamma(\kappa)}{2\pi^{\kappa+1} r_x^{2\kappa} \Gamma(2\kappa)} \sum_{j=0}^\infty (j+\kappa) \frac{\Gamma(j+2\kappa)}{\Gamma(j+1)} G_j(r,r) \,. \label{eqn: the tadpole}
\end{align}
We immediately notice two things: the tadpole only depends on the radius~$r$, and furthermore it is a sum over all partial waves~$G_j$, appropriately weighted by their degeneracy.

To study the divergent nature of the angular momentum sum, we must work with some approximate expression for the propagator for large~$j$. 
This we can obtain via a WKB expansion of the solution to Eq.~\eqref{eqn: 1 loop 2pi green eom}. 

In order to use the WKB method, we need to have an ordinary differential equation, meaning no convolution integral.
As we will see later in this section, the renormalised bubble can be written as the sum of a purely local and a purely non-local contribution
\begin{equation}
    \Sigma_j(r_x,r_y) = \: \Sigma_j^{\rm n.l.}(r_x,r_y) + \Sigma_j^{\rm loc} (r_x) \delta(r_x-r_y)\,,
\end{equation}
where~$\Sigma_j^{\rm n.l.}(r,r)=0$.
At the end of this section, we justify that~$\Sigma_j^{\rm n.l.}$ does not affect the ultraviolet (UV) properties of the theory. So, for the purpose of capturing the contributions causing UV divergences in the loops in a WKB propagator, we can drop it and, with it, the integral.
As for the local term, it only affects the UV behaviour of the propagator if
\begin{equation}\label{eqn: definition limit local part of the bubble}
    \lim_{j\to\infty} \Sigma_j^{\rm loc}(r) = \varsigma(r)\neq0\,,
\end{equation}
where this equation also serves as the definition of~$\varsigma(r)$.

With this, we solve Eq.~\eqref{eqn: 1 loop 2pi green eom} in the WKB approximation. In Appendix~\ref{app: WKB green}, we show the details of this calculation. Here, we only report the result
\begin{widetext}
\begin{equation}
    G^{\mathrm{WKB}}_j(r_x,r_y) = \: \frac{1}{2(j+\kappa)} \left( \frac{r_<}{r_>} \right)^{j+\kappa} \Bigg(1 - \frac{A (r_>,r_<)}{2(j+\kappa)} + \frac{A (r_>,r_<)^2}{8(j+\kappa)^2} - \frac{3 r_x^2 m_\varphi^2(r_x) - r_y^2 m_\varphi(r_y)^2}{4(j+\kappa)^2}  + \mathcal{O}(j^{-3}) \Bigg)\,,\label{eqn: WKB green full}
\end{equation}
\end{widetext}
where ${r_> = \max (r_x,r_y) }$, and $r_<= \min (r_x,r_y)$. 
We have defined the effective mass~${m_\varphi^2 = U''(\varphi) + \Pi_\varphi+\varsigma_\varphi}$, recalling the definition of~$\varsigma_\varphi$ in Eq.~\eqref{eqn: definition limit local part of the bubble}.
In Eq.~\eqref{eqn: WKB green full} we have introduced the function
\begin{equation}
    A (r_>,r_<) = \: \int_{r_<}^{r_>} \d \tilde r \, \tilde r \, m_\varphi^2 (\tilde r) \,. 
\end{equation}
In the coincident limit~$r_x=r_y$ we have~$A\equiv0$, and Eq.~\eqref{eqn: WKB green full} becomes more compact,
\begin{equation}
    \scalebox{0.9}{
    $\displaystyle 
    G^{\mathrm{WKB}}_j (r,r) = \: \frac{1}{2(j+\kappa)} \left( 1 - \frac{r^2 m_\varphi^2(r)}{2(j+\kappa)^2} + \mathcal{O}(j^{-4})\right)
    $}
    \,.
\end{equation}
By plugging this into Eq.~\eqref{eqn: the tadpole} we can isolate the divergent behaviour of the tadpole in the angular momentum representation
\begin{align}
    \Pi_\varphi^{\mathrm{UV}} (x) \sim \: & \frac{1}{2} \sum_j^\infty (j+\kappa)^{2\kappa-1}  \notag \\
    & \hspace{-4em}\times 
    \left( 1 - \frac{3r^2m_\varphi^2(r) +  \kappa(2\kappa-1) (\kappa-1)}{6(j+\kappa)^2} + \mathcal{O}(j^{-4}) \right)
    \,,\label{eqn: tadpole uv}
\end{align}
where we have also expanded the $\Gamma$-functions for large~$j$.
For~${d=4}$, namely~${\kappa=1}$, the angular momentum sum in Eq.~\eqref{eqn: tadpole uv} is both quadratically and logarithmically divergent, exactly as we would find for the tadpole in a homogeneous background in momentum space. For lower dimensions (lower~$\kappa$), the situation improves.
Higher order terms in the sum have been omitted as these are convergent for~${d\leq4}$.

The strategy for renormalising the tadpole (and any other diagram in the angular momentum representation) follows the BPHZ subtraction scheme: we subtract from the summand its WKB expression for large angular momentum~$j$, thus making the sum finite; then we add back the WKB expression, for which the sum can now be evaluated analytically in dimensional regularisation; finally we use counter-terms to subtract the divergences in the regularisation parameter.
For the last step, we use~$\overline{\mathrm{MS}}$ counter-terms.
In equations
\begin{equation}
    \Pi_\varphi(x) = \: \left[ \Pi_\varphi(x) - \Pi_\varphi^{\mathrm{UV}}\right]_d + \lim_{\epsilon\to0} \left[ \Pi_\varphi^{\mathrm{UV}} + \Delta_{\mathrm{c.t.}} \right]_{d-2\epsilon}\,.
    \label{eqn: BPHZ subtraction tadpole}
\end{equation}
Note that the more orders in the WKB expansion we include inside~$\Pi_\varphi^{\mathrm{UV}}$, the better we resolve the tadpole analytically through a gradient expansion. However, for the sake of simplicity, we include as many orders in the WKB expansion as is necessary to capture the divergent behaviour and not more.

Regularising divergences by subtracting the correlator computed within the WKB approximation is a way of implementing renormalisation in the 2PI formalism.
Differently from the usual perturbation theory, the 2PI propagators used in the loops are dressed propagators and, therefore, induce sub-divergences that the subtraction procedure addresses naturally, as we will further elaborate at the end of this section.

To compute the finite part of the renormalised sum in the last term of Eq.~\eqref{eqn: BPHZ subtraction tadpole} we must now focus on a specific value for~$d$. We make all of our explicit calculations for~${d=4}$. In Section~\ref{sec: parametric example d=2}, we present the same calculations for~${d=2}$, which are easier to do.
In~${d=4-2\epsilon}$, the WKB tadpole sum can be computed explicitly and reads 
\begin{widetext}
\begin{equation}
    \Pi_\varphi^{\mathrm{UV}} (r) = \: \frac{\lambda}{4} \frac{(\pi \tilde\mu^2 r_x^2)^\epsilon \Gamma(1-\epsilon)}{2\pi^2 r_x^2 \Gamma(2-2\epsilon)} \sum_{j=0}^\infty (j+1-\epsilon)^{1-2\epsilon} \left( 1 - \frac{3r^2 m_\varphi^2(r) - \epsilon}{6(j+1-\epsilon)^2} \right) = \: - \frac{\lambda m_\varphi^2(r)}{32\pi^2} \left(\frac{1}{\epsilon} + \log \mu^2 r^2 \right) \,,
\end{equation}
\end{widetext}
where we have introduced the renormalisation scale~${\mu^2 = e^{2+\gamma_E} \pi \tilde \mu^2}$ to impose the correct dimensionality of~$\Pi$.

For consistency, we can check explicitly that the divergence we get is the same as we would find when working in the trivial background in the conventional momentum representation.
In Fourier space, we can write the tree-level false vacuum Green's function 
\begin{equation}
    G_{\mathrm{FV}}(x,y) = \: \int \frac{\d^4 k}{(2\pi)^4} e^{- i k (x-y)} \frac{1}{k^2 + m^2} \,.
\end{equation}
Then the tadpole in dimensional regularisation reads
\begin{align}
    \Pi_{\varphi_{\mathrm{FV}}} = \: &  \frac{\lambda}{2} G(x,x) = \: \mu^{4-d} \int \frac{\d^d k}{(2\pi)^d} \frac{1}{k^2 + m^2} \notag \\
    = \: & - \frac{\lambda m^2}{32\pi^2 \epsilon}\,,
\end{align}
and we see the same pole as found in the bounce background at leading order in the coupling, namely for~$m_\varphi^2(r) = m^2$.
The divergence can then be renormalised by the usual counter-terms.
When using the full mass~$m_\varphi^2(r)$, there are additional divergences coming from sub-diagrams, as explained at the end of Section~\ref{subsec: the bubble}. These are then renormalised by higher-order counter-terms, which can be absorbed by trading~$m^2$ for~$m_\varphi^2(r)$ in the leading order ones.

The $\overline{\mathrm{MS}}$-renormalised tadpole, i.e. the explicit result for Eq.~\eqref{eqn: BPHZ subtraction tadpole} in this scheme, in four dimensions is
\begin{align}
    \Pi_\varphi^{\overline{\mathrm{MS}}} (r) = \: & \frac{\lambda}{4\pi^2r^2} \sum_{j=0}^\infty (j+1)^2 \left( G_j(r,r) - G_j^{\mathrm{WKB}} (r,r) \right) \notag \\
    & - \frac{\lambda m_\varphi^2(r)}{32\pi^2} \log \mu^2 r^2 \,,
\end{align}
where once again we retain in the WKB Green's function enough orders to leave the sum finite
\begin{equation}
    G_j^{\mathrm{WKB}} = \: \frac{1}{2(j+1)} \left( 1 - \frac{r^2m_\varphi^2(r)}{2(j+1)^2} \right) \,.
\end{equation}
Now, let us turn our attention to what renormalisation scheme is the most suitable for our purposes.
Above, we have chosen the $\overline{\mathrm{MS}}$-scheme for the sake of simplicity. Eventually, we want to relate the decay rate to physical quantities. Thus, we choose to work in the false-vacuum renormalisation scheme, which we define as the scheme where all corrections to the couplings vanish in the false vacuum background, i.e.
\begin{subequations}
\begin{align}
    U_{\mathrm{eff}}(\varphi_{\mathrm{FV}}) = \: & 0 \,, \\
    \frac{\partial U_{\mathrm{eff}} (\varphi)}{\partial \varphi} \Big\rvert_{\varphi = \varphi_{\mathrm{FV}}} = \: & 0 \,, \\
    \frac{\partial^2 U_{\mathrm{eff}} (\varphi)}{\partial \varphi^2} \Big\rvert_{\varphi = \varphi_{\mathrm{FV}}} = \: & m^2 \,, \label{eqn: mass false-vacuum condition} \\ 
    \frac{\partial^3 U_{\mathrm{eff}} (\varphi)}{\partial \varphi^3} \Big\rvert_{\varphi = \varphi_{\mathrm{FV}}} = \: & g \,, \\
    \frac{\partial^4 U_{\mathrm{eff}} (\varphi)}{\partial \varphi^4} \Big\rvert_{\varphi = \varphi_{\mathrm{FV}}} = \: & \lambda \,,
\end{align}
\end{subequations}
where~$U_{\mathrm{eff}}(\varphi)$ is the effective potential computed in the \emph{homogeneous} background~$\varphi$.
These renormalisation conditions ensure that~$\varphi_{\mathrm{FV}}$ remains a local minimum of the effective potential.
These conditions uniquely fix the counter-terms, and therefore the subtraction procedure that we must follow for each quantity that needs to be renormalised.

Going back to the tadpole, we can express the false-vacuum result by subtracting the~$\overline{\mathrm{MS}}$ tadpole computed in the false vacuum background
\begin{align}
    \Pi_\varphi^{\mathrm{OS}} (r) = \: & \Pi_\varphi^{\overline{\mathrm{MS}}} (r) - \Pi_{\varphi_{\mathrm{FV}}}^{\overline{\mathrm{MS}}} (r) \notag \\
    = \: & \frac{\lambda}{4\pi^2r^2} \sum_{j=0}^\infty (j+1)^2 \Bigg[ G_{\varphi,j}(r,r) - G_{{\varphi_{\mathrm{FV}}},j} (r,r) \notag \\
    & \qquad \qquad \qquad \qquad + \frac{r^2 ( m_\varphi^2(r) - m^2)}{4(j+1)^3} \Bigg] \notag \\
    & - \frac{\lambda(m_\varphi^2(r) - m^2)}{32\pi^2} \log \mu^2r^2 \,, \label{eqn: false-vacuum tadpole}
\end{align}
where the subscripts indicate in what background each Green's function is to be computed.
Equation~\eqref{eqn: false-vacuum tadpole} makes it explicit that the tadpole gives no correction to the mass in the false vacuum background, namely that~${\Pi_{\varphi_{\mathrm{FV}}}^{\mathrm{OS}} = 0}$.

\subsection{The bubble}\label{subsec: the bubble}
The bubble receives contributions both from the cubic and from the quartic vertices, as shown diagrammatically in Eq.~\eqref{eqn: diagrammatic self-energy}.
It reads
\begin{align}
    \Sigma_\varphi(x,y) = \: & \Big( \frac{g^2}{2} - \frac{g\lambda}{4} ( \varphi(r_x) + \varphi(r_y)) \notag \\
    & + \frac{\lambda^2}{2} \varphi(r_x) \varphi(r_y) \Big) G(x,y)^2 \,. \label{eqn: bubble position space}
\end{align}
Our first goal is to rewrite this in the form of Eq.~\eqref{eqn: ansatz bubble factorisation}, thus identifying~$\Sigma_{\varphi,j}$.
This is easier said than done.
\paragraph{Linearising the bubble.}
The vertices are already functions solely of the radii~$r_x$ and~$r_y$, and no further manipulation is needed for it.
Let us then focus on the square of the Green's function, for which we must carry out all of the manipulations
\begin{widetext}
\begin{align}
    G(x,y)^2 = \: & \left(\frac{\Gamma(\kappa)}{2\pi^{\kappa+1} (r_xr_y)^\kappa} \right)^2 \sum_{j,j'=0}^\infty (j+\kappa) (j'+\kappa) C_j^\kappa (\cos\theta) C_{j'}^\kappa (\cos\theta) G_j(r_x,r_y) G_{j'}(r_x,r_y) \notag \\
    \overset{!}{=} \: & \frac{\Gamma(\kappa)}{2\pi^{\kappa+1} (r_xr_y)^\kappa} \sum_{j=0}^\infty (j+\kappa) C_j^\kappa (\cos\theta) \mathbf{S}_j(r_x,r_y) \,, \label{eqn: definition linear green squared}
\end{align}
having defined the angle~$\theta$ by ${x\cdot y = |x||y| \cos\theta}$, and where~$\mathbf{S}_j$ is to be determined.
We need the following linearisation formula for the Gegenbauer polynomials
\begin{equation}\label{eqn: Gegenbauer linearisation formula}
    C_j^\kappa(\cos\theta) C_{j'}^\kappa(\cos\theta) = \: \sum_{p=0}^{\min (j,j')} \gamma_{jj'p}^{(\kappa)} C_{j+j'-2p}^\kappa(\cos\theta)\,,
\end{equation}
which we take from Ref.~\cite{doi:10.1137/1.9781611970470.ch5}. A proof of it can be found in the references therein.
The linearisation coefficients~$\gamma_{jj'p}^{(\kappa)}$ are explicitly given in Eq.~\eqref{eqn: linearisation coefficients}.
After a long derivation, which we present in all of its details in Appendix~\ref{app: linearising square green}, we find 
\begin{align}
    \mathbf{S}_j(r_x,r_y) = \: & \frac{\Gamma(\kappa)}{2\pi^{\kappa+1} (r_x r_y)^\kappa} \frac{1}{j + \kappa} \left[ 2\sum_{q=\left\lfloor \frac{j+2}{2} \right\rfloor}^j \sum_{\ell=q}^\infty \alpha^{\varphi,(\kappa)}_{\ell,\ell-2q+j,\ell-q} (r_x,r_y) + \frac{1+(-1)^j}{2} \sum_{\ell=\frac{j}{2}}^\infty \alpha^{\varphi,(\kappa)}_{\ell,\ell,\ell-\frac{j}{2}} (r_x,r_y) \right]\,, \label{eqn: linearised green squared}
\end{align}
\end{widetext}
where we have introduced the shorthand notation
\begin{equation}\label{eqn: definition alphas}
    \alpha^{\varphi,(\kappa)}_{\ell,\ell',q} (r_x,r_y) = \: (\ell+\kappa)(\ell'+\kappa) \gamma^{(\kappa)}_{\ell \ell' q} G_{\varphi,\ell} (r_x,r_y) G_{\varphi,\ell'} (r_x,r_y) \,.
\end{equation}
Note that the sum over~$q$ goes from~${\left\lfloor \frac{j+2}{2} \right\rfloor}$ to~$j$, implying that when~$j=0$, there is nothing to sum onto, and this term vanishes.
While the expression~(\ref{eqn: linearised green squared}) might appear a bit daunting, we recognise a structure that is familiar from (Euclidean) momentum space, where the bubble in a homogeneous background takes the schematic form
\begin{equation}
    \mathbf{S} (p) = \: \int \frac{\d^d k}{(2\pi)^d} \, \frac{1}{(p-k)^2 + m^2 } \, \frac{1}{k^2+m^2} \,,
\end{equation}
with~$p$ the external momentum and~$k$ the momentum running in the loop.
Analogously, in Eq.~\eqref{eqn: linearised green squared}~$j$ is the external angular momentum, while~$\ell$ is the angular momentum running in the loop. We have an additional label~$q$ because of the more complicated linearisation formula for the basis functions in this representation.
For comparison, in momentum space, the analogous linearisation formula is simply
\begin{equation}
    e^{-i p_1\cdot (x-y)} \, e^{-i p_2 \cdot (x-y)} = \: e^{-i (p_1+p_2) \cdot (x-y)} \,.
\end{equation}
\paragraph{Extracting divergences.}
Now, we can turn to the analysis of the UV properties of the "loop sum", encoded in the large~$\ell$ behaviour of the summand.
Once again, we need the WKB-expanded Green's function from Eq.~\eqref{eqn: WKB green full}. From considerations in momentum space, we expect the loop sum in Eq.~\eqref{eqn: linearised green squared} to be divergent only for~${d\geq4}$. In the following, we look at it for~${d=4-2\epsilon}$.
We only need to keep the leading term in the WKB Green's function in Eq.~\eqref{eqn: WKB green full} as will become clear \textit{a posteriori}.
Let us look at the possibly divergent sum in the first term of Eq.~\eqref{eqn: linearised green squared}. Expanding this summand for large loop angular momentum~$\ell$, we get
\begin{align}
    (\ell+\kappa) & (\ell-2q+j+\kappa) \gamma_{\ell,\ell-2q+j,\ell-q}^{(\kappa)} G_\ell G_{\ell-2q+j} \notag \\
    = \: & (\ell+\kappa)^{2(\kappa-1)} 
    \scalebox{0.8}{$\displaystyle \frac{(j+\kappa)\Gamma(j+1) \Gamma\left(j + \kappa - q\right) \Gamma\left(q + \kappa\right)}{\Gamma(\kappa)^2 \Gamma(j+2\kappa) \Gamma\left(j+1-q\right) \Gamma\left(q+1\right)} $} \notag \\
    & \times \frac{1}{4} \left(\frac{r_<}{r_>} \right)^{2\ell - 2q +j + 2\kappa} \left( 1 +\mathcal{O} \left(\frac{1}{\ell+\kappa}\right)\right)\,, \label{eqn: uv behaviour sum terms}
\end{align}
where we neglected higher orders in the expansion as these terms lead to convergent series in the distributional sense for all relevant values of $\kappa$, even in the coincident limit.

The first observation is that for~$r_x\neq r_y$, we have a geometric series with argument~${r_</r_> \leq 1}$.
As a consequence, the sum is clearly convergent except for the coincident limit when~${r_x=r_y}$. This is a good consistency check and implies that any divergence, if present, is local; namely we expect
\begin{equation}
    \mathbf{S}_j \rvert_{\mathrm{div.}} \simeq \frac{1}{\epsilon} \delta(r_x-r_y)\,.
\end{equation}

Next, we want to extract this behaviour explicitly.
Note that~${\mathbf{S}_j(r_x,r_y)}$, or equivalently~${\Sigma_{\varphi,j}(r_x,r_y)}$, will only appear as a distribution, namely inside a convolution, cf. Eq.~\eqref{eqn: 1 loop 2pi green eom}.
Then, when convoluting with a test function~$g(r_y)$, extracting the divergent part of the distribution amounts to the following procedure,
\begin{align}
    &\int_0^\infty \hspace{-0.7em} \d r_y \, r_y \mathbf{S}_j(r_x,r_y) \, g(r_y)\notag\\ =\: &  \int_0^\infty \hspace{-0.7em} \d r_y \, r_y \mathbf{S}_j(r_x,r_y) \left[ g(r_y) - g(r_x)\right] \notag \\
     +& g(r_x) \int_0^\infty \hspace{-0.7em} \d r_y \, r_y \mathbf{S}_j(r_x,r_y) \,,
\end{align}
or in the language of distributions,
\begin{equation}
    \mathbf{S}_j(r_x,r_y) = \: \left\lfloor \mathbf{S}_j(r_x,r_y) \right\rfloor_+ + \delta(r_x-r_y) \int_0^\infty \hspace{-0.7em} \d r_z \, r_z \mathbf{S}_j(r_x,r_z) \,.
\end{equation}
Here we are defining the~+ distribution by its action on a test function
\begin{align}
    &\int_0^\infty \hspace{-0.7em} \d r_y \, r_y \, \left\lfloor \mathbf{S}_j(r_x,r_y) \right\rfloor_+ g(r_y)\notag\\ = \:& \int_0^\infty \hspace{-0.7em} \d r_y \, r_y \, \mathbf{S}_j(r_x,r_y) \left[ g(r_y) - g(r_x)\right] \,.
\end{align}
We see that the bubble still contains a local part, and our task is now to identify and subtract its UV divergences.
We start by integrating the asymptotic term as expanded in Eq.~\eqref{eqn: uv behaviour sum terms}. Formally, we are exchanging the series~$\sum_{\ell = q}^\infty$ in Eq.~\eqref{eqn: linearised green squared} with the integral over the radius~$r_y$. This is allowed by dominated convergence, since for any~$\epsilon>0$ the series is convergent.
To do the integral we perform a useful change of variable
\begin{equation}
    \int_0^\infty \d r_y \, r_y f\left(\frac{r_<}{r_>}\right) = \: r_x^2 \int_0^1 \d z \, \left(z + \frac{1}{z^3} \right) f(z) \,,
\end{equation}
which in our case reads
\begin{equation}
    \int_0^\infty \d r_y \, r_y \left(\frac{r_<}{r_>} \right)^{2\ell - 2q + j + 2\kappa} \hspace{-2em}= \: r_x^2 \frac{\ell-q+\frac{j}{2}+\kappa}{\left(\ell-q+\frac{j}{2}+\kappa\right)^2-1} \,.
\end{equation}
Now, to compute the infinite sum, we split it into a finite term plus a term that diverges for~${\kappa=1}$,
\begin{align}
    &\sum_{\ell=j}^\infty  \frac{(\ell+\kappa)^{2(\kappa-1)} \left(\ell-q+\frac{j}{2}+\kappa\right)}{\left(\ell-q+\frac{j}{2}+\kappa\right)^2-1} \notag \\
    = \: & \sum_{\ell=j}^\infty \left( \frac{(\ell+\kappa)^{2(\kappa-1)} \left(\ell-q+\frac{j}{2}+\kappa\right)}{\left(\ell-q+\frac{j}{2}+\kappa\right)^2-1} - \frac{1}{\ell^{3-2\kappa}} \right)\notag\\ + & \sum_{\ell=j}^\infty \frac{1}{\ell^{3-2\kappa}} \,.
\end{align}
The first term is now a convergent series when~$\kappa=1$ and evaluates to
\begin{align}
    & \sum_{\ell=q}^\infty  \left(\frac{\ell-q+\frac{j}{2}+1}{(\ell-q+\frac{j}{2}+1)^2-1} - \frac{1}{\ell} \right) \notag\\
    = \: & \psi(q) - \frac{1}{2} \psi\left(\frac{j}{2}\right) - \frac{1}{2} \psi\left(2 + \frac{j}{2}\right)\,,
\end{align}
with~$\psi$ being the di-gamma function.
As for the divergent series, we expand it for small~$\epsilon$, where~${\kappa=1-\epsilon}$,
\begin{align}
    \sum_{\ell=q}^\infty \frac{1}{\ell^{3-2\kappa}} = \: & \zeta_{3-2\kappa}(q) = \: \frac{1}{2\epsilon} - \psi(q) + \mathcal{O}(\epsilon) \,.
\end{align}
As expected, we find a~$1/\epsilon$ pole.
Putting everything together, we can re-write the first term of Eq.~\eqref{eqn: linearised green squared} for~$\kappa=1-\epsilon$ and for odd~$j$
\begin{widetext}
\begin{align}
    & \mathbf{S}_j\rvert_{\mathrm{odd}} (r_x,r_y) = \: \left\lfloor \frac{1}{2\pi^2 r_xr_y} \frac{2}{j+1} \sum_{q=\frac{j+1}{2}}^{j} \sum_{\ell=q}^\infty 
    \alpha^{\varphi,(1)}_{\ell,\ell-2q+j,\ell-q} (r_x,r_y)
    \right\rfloor_+ \notag \\
    & + \delta(r_x-r_y) \frac{1}{2\pi^2 r_x^2} \frac{2}{j+1} \sum_{q=\frac{j+1}{2}}^{j} \sum_{\ell=q}^\infty 
     \int_0^\infty \d r_z\, r_z 
    \left( 
    \alpha^{\varphi,(1)}_{\ell,\ell-2q+j,\ell-q}(r_x,r_z) 
    - \frac{1}{4} \left(\frac{r_<}{r_>}\right)^{2(\ell-q+1) + j} \right)  \notag \\
    & + \delta(r_x-r_y) \frac{1}{16\pi^2} \Bigg( \frac{1}{\epsilon} - 3  + \log \pi \mu^2 r_x^2 - \frac{2}{j+2} -\frac{2}{j} + 2H_{j+1} + 2 \psi\left(\frac{j}{2}\right) + \psi\left(\frac{j+3}{2}\right) \Bigg) \,,
\end{align}
where~$H_n$ is the~$n^{\mathrm{th}}$ harmonic number, and~$\mu$ is the renormalisation scale already introduced in the renormalisation of the tadpole.

Once again, let us check that the divergence is indeed what we expect from calculations in the trivial background.
First of all, we must still carry out the sum over~$j$ to find the pole of~$G(x,y)^2$. The remaining contributions from Eq.~\eqref{eqn: linearised green squared} lead to the same divergent term.
Then, summing over~$j$ amounts to summing over the Gegenbauer polynomials, times a constant, which yields back the angular~$\delta$-function. Together with the radial~$\delta$-function above, we find
\begin{equation}
    G(x,y)^2 \rvert_{\mathrm{pole}} = \: \frac{1}{16\pi^2 \epsilon} \delta^{(4)}(x-y) \,.
\end{equation}
Now, in the trivial background we can work in momentum representation to find
\begin{align}
    G(x,y)^2 = \: & \int \frac{\d^4 p}{(2\pi)^4} e^{-ip(x-y)}\int \frac{\d^4k}{(2\pi)^4} e^{-ik(x-y)} \frac{1}{p^2+m^2} \frac{1}{k^2+m^2} \notag \\
    = \: & \int\frac{\d^4 p}{(2\pi)^4} e^{-ip(x-y)} \mu^{4-d} \int \frac{\d^dk}{(2\pi)^d} \frac{1}{(p-k)^2+m^2} \frac{1}{k^2+m^2} = \: \frac{1}{16\pi^2 \epsilon} \delta^{(4)}(x-y) + \mathrm{finite} \,,
\end{align}
which matches exactly the UV pole we have found in the bounce background.
Once again, this can be renormalised via usual counter-terms. No additional divergences are introduced by sub-diagrams, as explained at the end of this section.

\paragraph{Renormalisation.}
To renormalise, we impose the false-vacuum condition~\eqref{eqn: mass false-vacuum condition}, namely that the local part vanishes in the false vacuum background.
The contributions from even~$j$ and from the last term in Eq.~\eqref{eqn: linearised green squared} follow on analogous lines as the derivation we have just made. While the result of the regularised sum is slightly different, it maintains the property of being independent of the background field~$\varphi$.
Thus, we finally find the false-vacuum renormalised bubble self-energy
\begin{align}
    \mathbf{S}_j^{\mathrm{OS}} (r_x,r_y) = \: & \left\lfloor \frac{1}{2\pi^2 r_xr_y} \frac{1}{j+1} \sigma_j^{\varphi} (r_x,r_y) \right\rfloor_+ + \delta(r_x-r_y) \frac{1}{2\pi^2 r_x^2} \frac{1}{j+1} \int_0^\infty \d r_z \, r_z \left[ \sigma_j^{\varphi}(r_x,r_z) - \sigma_j^{\varphi_{\mathrm{FV}}} (r_x,r_z) \right]\,, \label{eqn: renormalised bubble 4d}
\end{align}
\end{widetext}
where we have introduced
\begin{align}
    \sigma_j^{\varphi}(r_x,r_y) = \: & 2 \sum_{q=\left\lfloor \frac{j+2}{2} \right\rfloor}^j \sum_{\ell=q}^\infty \alpha^{\varphi,(1)}_{\ell,\ell-2q+j,\ell-q} (r_x,r_y) \notag \\
    & + \frac{1+(-1)^j}{2} \sum_{\ell=\frac{j}{2}}^\infty \alpha^{\varphi,(1)}_{\ell,\ell,\ell-\frac{j}{2}} (r_x,r_y) \,.
\end{align}
In the following, we will avoid using the superscript OS, since all renormalised quantities are understood to be in the false-vacuum scheme as defined above.
In subtracting the local contribution in the false vacuum background, all local terms that do not depend on the background have dropped out.
In four dimensions we find a finite local term contributing to the bubble self-energy
\begin{equation}\label{eqn: bubble non-local plus local}
    \Sigma_j(r_x,r_y) = \: \Sigma_j^{\rm n.l.}(r_x,r_y) + \Sigma_j^{\rm loc} (r_x) \delta(r_x-r_y)\,,
\end{equation}
and
\begin{align}
    \Sigma_j^{\mathrm{loc}} (r_x) = \: & \left( \frac{g^2}{2} - \frac{g\lambda}{2} \varphi(r_x) + \frac{\lambda^2}{2} \varphi(r_x)^2 \right) \frac{1}{2\pi^2 r_x^2} \frac{1}{j+1} \notag \\
    & \hspace{-2em}\times  \int_0^\infty \hspace{-0.6em} \d r_z \, r_z \left[ \sigma_j^{\varphi}(r_x,r_z) - \sigma_j^{\varphi_{\mathrm{FV}}} (r_x,r_z) \right] \,.
\end{align}

\paragraph{On the WKB expansion} 
When obtaining the WKB-expanded Green's function in Eq.~\eqref{eqn: WKB green full}, we have dropped terms in Eq.~\eqref{eqn: 1 loop 2pi green eom}, and in particular (part of) the convolution integral by arguing that these do not affect the UV behaviour of the theory.
Here, we work out the details of this argument. We focus on~$d=4$, and we comment on lower dimensional cases at the end of this discussion on the WKB expansion.
As we will argue shortly, to assess whether a term is relevant in the UV or not when solving Eq.\eqref{eqn: 1 loop 2pi green eom} in the WKB approximation, its behaviour for large~$j$ is crucial.
The tadpole is trivially independent of~$j$, so we focus now on the bubble~$\Sigma_j$ as given in Eq.~\eqref{eqn: bubble non-local plus local} in terms of a purely non-local and purely local term.
The non-local part is always exponentially suppressed because of the presence of the term~$(r_</r_>)^j$, 
as can be seen from the WKB expansion of the Green's function in Eq.~\eqref{eqn: WKB green full}, and also from the behaviour of the terms in the sum as given in Eq.~\eqref{eqn: definition alphas} and Eq.~\eqref{eqn: uv behaviour sum terms}.
The purely local part, on the other hand, does not need to vanish. In particular, we can have a non vanishing~$\varsigma(r)$ as defined in Eq.~\eqref{eqn: definition limit local part of the bubble}.
Then, the left-over~$j$-dependent part vanishes, namely~${\lim_{j\to\infty}\Sigma_j^{\rm loc}(r) - \varsigma(r) = 0}$ and is irrelevant for computing the WKB Green's function.

Now, take the tree-level Green's functions, i.e. the solutions to Eq.~\eqref{eqn: 1 loop 2pi green eom} where we set both the tadpole and the bubble to zero.
Then, we can compute the renormalised tadpole and bubble as defined in Eqs.~\eqref{eqn: false-vacuum tadpole} and~\eqref{eqn: renormalised bubble 4d}, respectively, where the effective mass~$m_\varphi^2$ in the WKB Green's function coincides with the tree-level mass~$U''(\varphi)$.
We then plug the computed self-energy back into Eq.~\eqref{eqn: 1 loop 2pi green eom}, and compute again the Green's function keeping fixed the tadpole and the bubble.
When using this partial summation of the propagator to compute once again the tadpole and the bubble, we might be worried that new divergent diagrams arise that are not regularised by the WKB subtraction procedure.
Let us address this worry by first computing the bubble diagram using the partial summation of the propagator.
It is convenient to expand the latter in a geometric series, where the self-energy insertion enters as an interaction. In the following, thick lines represent dressed propagators, thin ones are the tree-level propagators, and squares represent insertions of the renormalised self-energy~$\mathbf{\Sigma}$ computed with tree-level propagators.
Neglecting vertex rules, the bubble with dressed propagators can be expanded as
\begin{equation}
    \begin{tikzpicture}
    \begin{scope}[shift={(0,0)}]
    \fill (-0.5,0) circle (0.1);
    \draw[very thick] (0,0) circle (0.5);
    \fill (0.5,0) circle (0.1);
    \end{scope}
    \draw (1,0) node {=};
    \begin{scope}[shift={(2,0)}]
    \fill (-0.5,0) circle (0.1);
    \draw (0,0) circle (0.5);
    \fill (0.5,0) circle (0.1);
    \end{scope}
    \draw (3,0) node {+};
    \begin{scope}[shift={(4,0)}]
    \fill (-0.5,0) circle (0.1);
    \draw (0,0) circle (0.5);
    \fill (0.5,0) circle (0.1);
    \fill (0+0.15,0.5+0.15) rectangle (0-0.15,0.5-0.15);
    \end{scope}
    \draw (2,-1.5) node {+};
    \begin{scope}[shift={(3,-1.5)}]
    \fill (-0.5,0) circle (0.1);
    \draw (-0.5,0) arc(180:90:0.5);
    \fill (0+0.15,0.5+0.15) rectangle (0-0.15,0.5-0.15);
    \draw (0,0.5) -- (0.5,0.5);
    \fill (0.5+0.15,0.5+0.15) rectangle (0.5-0.15,0.5-0.15);
    \draw (0.5,0.5) arc(90:0:0.5);
    \draw (-0.5,0) arc(180:270:0.5);
    \draw (0,-0.5) -- (0.5,-0.5);
    \draw (0.5,-0.5) arc(270:360:0.5);
    \fill (1,0) circle (0.1);
    \end{scope}
    \draw (5,-1.5) node {+ ...};
    \end{tikzpicture}
    \label{eqn: diagrammatic resummed bubble}
\end{equation}
The first term in the series is divergent and regularised by subtracting the same diagram computed with WKB propagators with tree-level mass~$U''(\varphi)$.
The second term reads schematically
\begin{equation}
    \begin{tikzpicture}
    \begin{scope}[shift={(0,0)}]
    \fill (-0.5,0) circle (0.1);
    \draw (0,0) circle (0.5);
    \fill (0.5,0) circle (0.1);
    \fill (0+0.15,0.5+0.15) rectangle (0-0.15,0.5-0.15);
    \end{scope}
    \draw (2,0) node {$\approx \sum_j j^2 G_j^3 \mathbf{\Sigma_j}$};
    \end{tikzpicture}
    \,.\label{eqn: schematic bubble with one insertion}
\end{equation}
The asymptotic behaviour of the Green's function for large~$j$ can be read off from Eq.~\eqref{eqn: WKB green full}, namely~$G_j\sim j^{-1}$.
Therefore, we can immediately conclude that any term in~$\mathbf{\Sigma}_j$ that decreases for large~$j$ will not give any additional divergence in the diagram~\eqref{eqn: schematic bubble with one insertion}.
However,~$\mathbf{\Sigma}_j$ contains the tadpole~$\Pi_\varphi$ and possibly the finite limit of the purely local part of the bubble~$\varsigma_\varphi$, both of which are~$j$-independent.
These induce an additional logarithmic divergence, and to get rid of it, we must then subtract from Eq.~\eqref{eqn: diagrammatic resummed bubble} the bubble diagram computed using WKB propagators evaluated with an effective mass~${m_\varphi^2 = U''(\varphi) + \Pi_\varphi + \varsigma_\varphi}$.
The subsequent terms in Eq.~\eqref{eqn: diagrammatic resummed bubble} are all finite, by a similar powercounting argument.

We can now look at the tadpole computed with partially summed propagators
\begin{equation} \label{eqn: diagrammatic dressed tadpole}
    \begin{tikzpicture}
    \begin{scope}[shift={(0,0)}]
    \draw[very thick] (0,0.5) circle (0.5);
    \fill (0,0) circle (0.1);
    \end{scope}
    \draw (1,0) node {=};
    \begin{scope}[shift={(2,0)}]
    \draw (0,0.5) circle (0.5);
    \fill (0,0) circle (0.1);
    \end{scope} 
    \draw (3,0) node {+};
    \begin{scope}[shift={(4,0)}]
    \draw (0,0.5) circle (0.5);
    \fill (0,0) circle (0.1);
    \fill (0+0.15,1+0.15) rectangle (0-0.15,1-0.15);
    \end{scope} 
    \draw (5,0) node {+ ...};
    \end{tikzpicture}
\end{equation}
Again, the first term is the loop diagram computed with tree-level propagators and is regularised by subtraction of the tadpole computed using WKB propagators with the tree-level mass.
Let us look at the second term
\begin{equation}
    \begin{tikzpicture}
    \begin{scope}[shift={(0,0)}]
    \draw (0,0.5) circle (0.5);
    \fill (0,0) circle (0.1);
    \fill (0+0.15,1+0.15) rectangle (0-0.15,1-0.15);
    \end{scope} 
    \draw (2,0) node {$\approx \sum_j j^2 G_j^2 \mathbf{\Sigma_j}$};
    \end{tikzpicture}
    \,.
\end{equation}
Once again, only the~$j$-independent contribution in~$\mathbf{\Sigma}_j$, namely the tadpole~$\Pi_\varphi$ and possibly the limit of the purely local contribution to the bubble~$\varsigma_\varphi$, induces additional divergences, while the purely non-local part of the bubble~$\Sigma^{\rm n.l.}_\varphi$ does not.

We conclude that to regularise all divergences arising in the loop diagrams computed with dressed propagators, as they appear on the left-hand side of Eqs.~\eqref{eqn: diagrammatic resummed bubble} and~\eqref{eqn: diagrammatic dressed tadpole}, it is sufficient to subtract the diagrams computed with the WKB Green's functions using the effective mass
\begin{equation}
    m_\varphi^2(r) = \: U''(\varphi(r)) + \Pi_\varphi(r) + \varsigma_\varphi(r)\,.
\end{equation}
We have argued this for the first step of the self-consistent procedure, however it is trivial to extend this argument by induction to each step of the iteration procedure.

Some final comments on lower dimensions: The ``measure'' of the sum over the angular momentum is~$j$ to some power lower than two, implying that the convergence properties are better.
In particular, for~$d=2$, the measure is just~1, and not even the tadpole induces any new divergence, meaning that at each step of the self-consistent procedure, it is always enough to use WKB propagators computed with the tree-level mass~$m^2$ for the subtraction procedure.

\section{A parametric example for $d=2$}\label{sec: parametric example d=2}

\subsection{The model}
In this section, we present some numerical results for the self-consistent procedure of finding the bounce and Green's functions in two dimensions. 
We choose to work in two dimensions because of the better convergence properties of the algorithm. All qualitative features of our method are already present here.

Such numerical approaches have already been attempted in previous work. To the best of our knowledge, the first examples of employing the 2PI effective action formalism to false vacuum decay are reported in Refs.~\cite{Bergner:2003id,Baacke:2004xk}, where the authors study the backreaction of the propagator onto the bounce equation in two dimensions and in the Hartree approximation, namely neglecting the bubble contribution. In Ref.~\cite{Baacke:2006kv}, the same approximation is used to obtain numerical results in four dimensions.
The Hartree approximation introduces explicit breaking of translational symmetry, as observed already in Ref.~\cite{Bergner:2003id}, thus making the translational modes massive. As commented in Section~\ref{sec: decay rate review}, this is a known feature of effective actions: while the full effective action preserves the same symmetries of the quantum theory, some truncations of it might break those symmetries~\cite{Pilaftsis:2013xna,Garbrecht:2015cla}.

To address the explicit breaking of translational symmetry, the authors of Ref.~\cite{Garbrecht:2018rqx} go beyond the Hartree approximation by introducing an additional term in the equation for the propagator, which explicitly restores the translational symmetry and thus the masslessness of the translational modes.
In the present work instead, we focus on including the non-local term in the scalar self-energy. In this section, we demonstrate numerically the failure of the Hartree approximation for some region of the parameter space, where the bubble contribution becomes as relevant as the tadpole one.

For the remainder of this section, we apply a convenient redefinition of the field and the couplings as appearing in the action~\eqref{eqn: action generic scalar theory} with the potential in Eq.~\eqref{eqn: generic potential},
\begin{align}
    g = \: & 3 m^{3-\frac{d}{2}} \sqrt{\beta} \,, \\
    \lambda = \: & 3 m^{4-d} \beta \alpha \,, \\
    \varphi = \: & \frac{m^{\frac{d}{2}-1}}{\sqrt{\beta}} \phi \,,
\end{align}
such that after introducing the dimensionless coordinates~${\tilde x = m x}$, the action~\eqref{eqn: action generic scalar theory} becomes
\begin{equation}
    S[\varphi] = \: \frac{1}{\beta} \tilde S[\phi] \,,
\end{equation}
where
\begin{equation}
    \tilde S[\phi] = \: \int \d^d \tilde x \, \left( \frac{1}{2} \partial_i\phi \partial_i\phi + W(\phi)\right) \,,
\end{equation}
and the potential reads
\begin{equation}
    W (\phi) = \: \frac{1}{2}\phi^2 - \frac{1}{2} \phi^3 + \frac{\alpha}{8}\phi^4 \,.\label{eqn: rescaled potential}
\end{equation}
In this convenient parametrisation, all of the classical dynamics is governed by the dimensionless coupling~$\alpha$. 
The dimensionless coefficient $\beta$ plays the same role as $\hbar$, and enters together with quantum corrections.
For~${0<\alpha<1}$, the potential~\eqref{eqn: rescaled potential} exhibits a false vacuum at~${\phi=\phi_{\mathrm{FV}}=0}$ and a true vacuum at
\begin{equation}
    \phi=\phi_{\mathrm{TV}}=\frac{9}{4\alpha} \left(1-\frac{8\alpha}{9} + \sqrt{1 - \frac{8\alpha}{9}}\right) \,.
\end{equation}
For~$\alpha\to1$ the two minima become degenerate and we approach the thin-wall limit. On the other hand, for smaller~$\alpha$ we move away from the thin-wall limit, all the way to~$\alpha=0$ for which the potential is unbounded from below. For larger $\alpha$, we also move away from the thin wall. Then the TV becomes the FV until there is only one minimum left for $\alpha>9/8$.

\subsection{Equations of motion}
Setting~$d=2$, or equivalently~$\kappa=0$, and using the parametrisation introduced above, Eq.~\eqref{eqn: 1 loop 2pi eom phi}---the equation of motion for the background---becomes
\begin{align}
    -\frac{1}{r} \frac{\d}{\d r} r \frac{\d}{\d r} \phi(r) + W'(\phi(r)) + \beta \Pi_\phi (r) \phi(r) = \: 0\,. \label{eqn: eom 2d}
\end{align}
Here and in the following, we drop the~$\sim$ in the rescaled coordinate~$\tilde r$, as only this rescaled coordinate appears and there is no room for confusion.
The equation for the propagator~\eqref{eqn: 1 loop 2pi green eom} now reads
\begin{align}
    &\left( - \frac{1}{r} \frac{\d}{\d r} r \frac{\d}{\d r} + \frac{j^2}{r^2} + W''(\phi(r)) + \beta \Pi_\phi(r) \right) G_j(r,r') \notag \\
    &\qquad\qquad + \beta \int_0^\infty \d r'' \, r'' \, \Sigma_{\phi,j}(r,r'') \, G_j(r'',r')  \notag \\
    & \qquad\quad = \: \frac{1}{r}\delta(r-r') - \delta_{j,1} \phi_{\mathrm{tr}}(r) \,\phi_{\mathrm{tr}} (r') \,. \label{eqn: green eq 2d}
\end{align}

The tadpole self-energy in~$d=2$ is only logarithmically divergent, namely the sum in Eq.~\eqref{eqn: the tadpole} can be regularised by simply subtracting the leading term in the WKB expansion in Eq.~\eqref{eqn: tadpole uv}. 
In particular, the leading term is fully independent of the background field, thus no finite part of the regularised sum is left after renormalisation,
\begin{equation}\label{eqn: tadpole 2d}
    \Pi_\phi(r) = \: \frac{3\alpha}{2\pi} \sum_{j=0}^\infty \left[ G_{\phi,j}(r,r) - G_{\phi_{\mathrm{FV}},j} (r,r) \right]\,.
\end{equation}
As for the bubble self-energy, it is not divergent at all, and it reads 
\begin{align}
    \Sigma_j (r_x,r_y) = \: & \frac{9}{2} \Big[ 1 - \frac{\alpha}{2} ( \phi(r_x) + \phi(r_y)) \notag \\
    & + \alpha^2 \phi(r_x) \phi(r_y) \Big] \, \mathbf{S}_j (r_x,r_y) \,,\label{eqn: bubble 2d}
\end{align}
where
\begin{align}
    2\pi \mathbf{S}_j 
    = \: & \left(1- \frac{1}{2} \delta_{j,0} \right) \sum_{\ell = \left\lceil \frac{j+1}{2} \right\rceil}^\infty G_\ell G_{|j-\ell|} \notag \\
    & + \frac{1}{2} G_0^2 + \frac{1+(-1)^j}{4} G_{\frac{j}{2}}^2 \,. \label{eqn: linearised green squared 2d}
\end{align}
A detailed derivation of how to take the~$\kappa\to0$ limit of Eq.~\eqref{eqn: linearised green squared} is provided in Appendix~\ref{app: bubble in d=2}.

\subsection{Tree level}
Having set up the equations to solve, namely Eq.~\eqref{eqn: eom 2d} for the background and Eq.~\eqref{eqn: green eq 2d} for the Green's function, we start by first solving them at tree-level, namely dropping the contribution of the self-energy. 
We obtain the tree-level bounce~$\phi^{(0)}_{b}$ by implementing a shooting algorithm to impose the boundary conditions that the background vanishes at~$r\to\infty$ and that the solution is regular at~$r=0$, namely~$\dot\phi^{(0)}_{b}(0)=0$.

Once we have the bounce, we find the zero and negative modes by solving the eigenvalue problem for the operator
\begin{equation}
    \mathcal{O}^{(0)}_{b,j} (r) = \: - \frac{1}{r} \frac{\d}{\d r} r \frac{\d}{\d r} + \frac{j^2}{r^2} + W''(\phi^{(0)}_{b}(r)) \,.
\end{equation}
This can be done numerically on \textit{Mathematica}, for example via the built-in function \texttt{NDEigensystem}.
For~$j=0$, we find a negative mode, while for~$j=1$, we find the expected zero mode, which coincides with~$\dot\phi_b^{(0)}$ up to normalisation.

At this point we turn to the Green's function.
It is convenient to first obtain the free Green's function in the false vacuum background, which is known for arbitrary dimension. In particular, we have
\begin{equation}
    \left[ - \frac{1}{r} \frac{\d}{\d r} r \frac{\d}{\d r} + \frac{(j+\kappa)^2}{r^2} + 1 \right] G^{(0)}_{\mathrm{FV},j}(r,r') = \: \frac{1}{r} \delta(r-r') \,,
\end{equation}
with the Green's function being
\begin{equation}
    G^{(0)}_{\mathrm{FV},j}(r,r') = \: I_{j+\kappa}(r_<) K_{j+\kappa}(r_>) \,,
\end{equation}
where~$I$ and~$K$ are modified Bessel functions of the first and of the second kind respectively.
Then, we express the Green's function in the bounce background as
\begin{equation}
    G^{(0)}_{b,j}(r,r') = \: N(r') I_{j+\kappa}(r_<) K_{j+\kappa}(r_>) f_{L,j}(r_<) f_{R,j}(r_>) \,,
\end{equation}
and solve numerically for~$f_{L,j}$ and~$f_{R,j}$, the left- and right-regular solutions of the homogeneous equation, respectively.
The normalisation~$N(r')$ is fixed by the pre-factor of the~$\delta$-function, in this case~$1/r$. It reads
\begin{align}
    & N(r) = \: f_{L,j}(r) f_{R,j}(r) \notag \\
    & \quad + \frac{I_{j+\kappa}(r) K_{j+\kappa}(r)}{\mathcal{W}[I_{j+\kappa}(r),K_{j+\kappa}(r)]} \mathcal{W} [f_{L,j}(r), f_{R,j}(r)] \,,
\end{align}
where~${\mathcal{W}[f,g]=fg'-f'g}$ is the Wronskian of~$f$ and~$g$.
More care must be taken for~$j=1$, where the Green's function equation~\eqref{eqn: green eq 2d} is non-homogeneous and we must subtract the zero-mode contribution.
The details of the subtraction procedure can be found in Ref.~\cite{Garbrecht:2018rqx}, and we briefly review them in Appendix~\ref{app: subtracting the zero-mode}.

We stop at a large enough~$j=j_{\rm max}$, matching the precision we require. 
The truncation is equivalent to taking the Green's function to be equal to the WKB approximated one for all~$j>j_{\rm max}$.
The error we make is a power of~$j_{\rm max}$, in particular
\begin{equation}
    G_{b,j}(r,r) - G_{b,j}^{\rm WKB}(r,r) \sim \frac{1}{j^3} \underset{j>j_{\rm max}}{<} \frac{1}{j_{\rm max}^3}\,.
\end{equation}
In our numerical example we take~$j_{\rm max}=20$, which ensures a precision better than~$10^{-3}$.

\subsection{Iterative procedure}
Having found the Green's function at tree-level, we use it to compute the tadpole and the bubble, then compute the bounce and the Green's function including quantum corrections by solving Eqs.~\eqref{eqn: eom 2d} and~\eqref{eqn: green eq 2d}. We then repeat the procedure until convergence, namely until neither the bounce nor the Green's function change significantly any more. 
For the purpose of better convergence, at each step of the iteration procedure we keep memory of the previous iteration. We do so, by only partially updating the tadpole and the bubble, namely if~$\widetilde{\Pi}_b^{(n)}$ is the tadpole computed with the Green's functions obtained at the~$n^{\mathrm{th}}$ iteration, we define the tadpole~$\Pi_b^{(n)}$, namely the function that enters the evaluation of the bounce and of the Green's function at the next iteration, as 
\begin{equation}
    \Pi_b^{(n)} = \: \zeta \widetilde{\Pi}_b^{(n)} + (1-\zeta) \Pi_b^{(n-1)} \,,
\end{equation}
where~$\zeta$ is a mixing parameter. The bubble is handled accordingly.
By only partially updating the value of the tadpole and the bubble, we prevent the solution from going astray and diverging far away from the solution we seek. This is of particular relevance when our desired solution is a saddle point in the functional space of the self-consistent procedure.
Empirically,~$\zeta=0.7$ is an optimal choice for our procedure.
A larger mixing parameter, in fact, renders the procedure more unstable so that we can stray away from the solution we seek very fast.
A smaller parameter, on the other hand, renders the convergence of the solution very slow and computationally expensive. This might even lead to satisfying the convergence criteria we set before having significantly moved away from the classical solution.

In studying Eq.~\eqref{eqn: green eq 2d}, we must address how we plan to solve an integro-differential equation.
A variety of methods are available, but mostly we might attempt two different strategies.
One straightforward idea would be to compute the integral appearing in the equation using a sensible ansatz for the Green's function, namely the one at the previous iteration to start with. This would result in a function of~$r$ and~$r'$ that plays the role of an inhomogeneity in the differential equation. We could then solve this ordinary differential equation self-consistently. The procedure is computationally very expensive, and given the complicated form of the bubble self-energy, it will be necessary to interpolate it as a function of~$r$ and~$r'$ if we hope to solve this in a reasonable amount of computing time.

One more sophisticated approach is to fit the bubble self-energy via a suitable basis of orthonormal functions, truncated at some high enough order, and then make an ansatz for the Green's function in terms of such basis functions. The equation can then be cast into a system of algebraic equations for the coefficients of the Green's function in the chosen basis. We could then solve for these coefficients self-consistently.

\begin{figure*}
    \centering
    \begin{subfigure}[b]{0.45\textwidth}
        \centering
        \includegraphics[width=\textwidth]{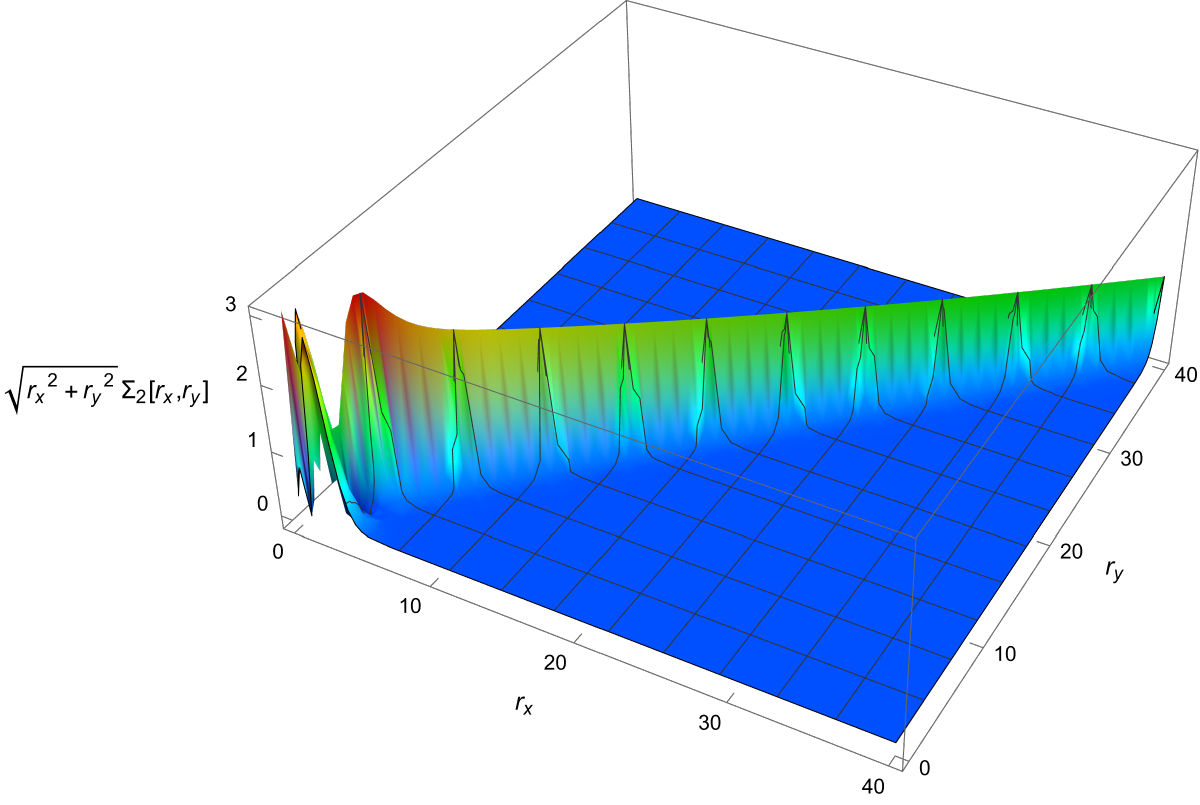}
        \caption{}
    \end{subfigure}
    \hfill
    \begin{subfigure}[b]{0.45\textwidth}
        \centering
        \includegraphics[width=\textwidth]{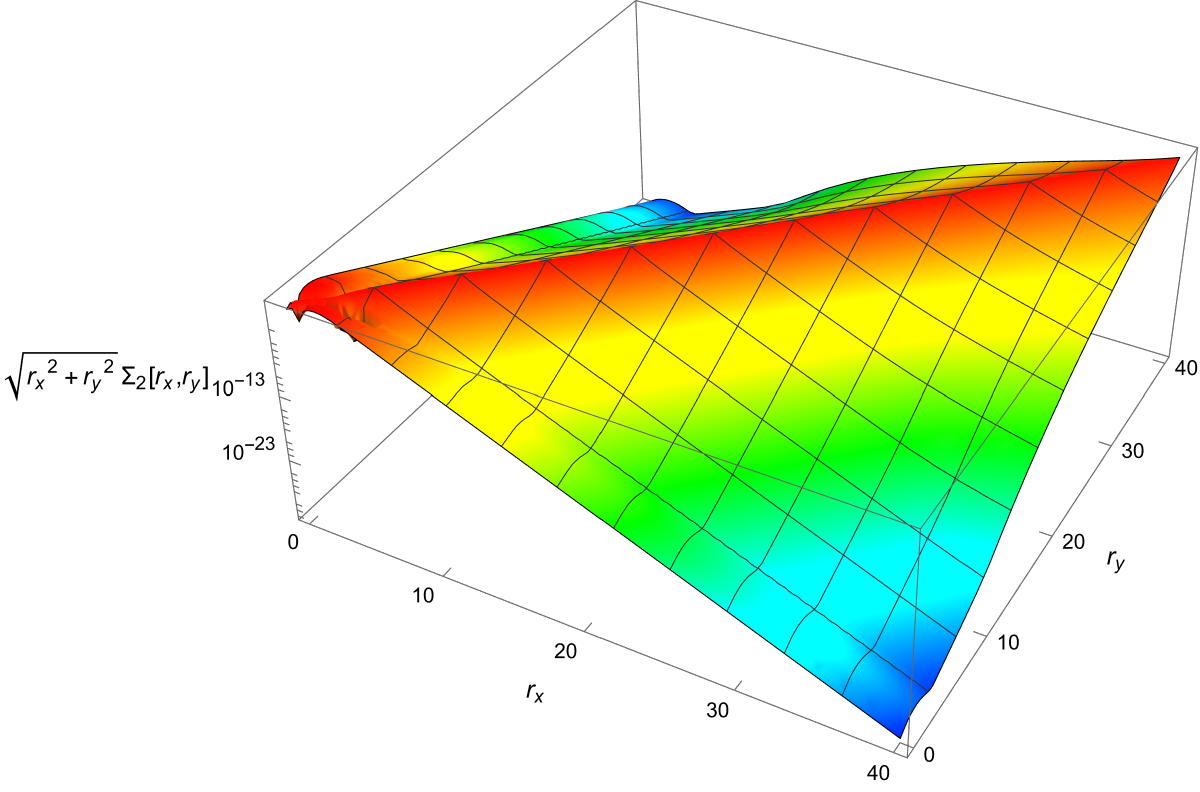}
        \caption{}
    \end{subfigure}
        \caption{Bubble self-energy for~$\alpha=0.7$, computed with tree-level propagators. Plotted in linear scale~((a) {\bf left}) and logarithmic scale ((b) {\bf right}). The linear decay away from the coincident diagonal in the logarithmic scale shows the exponential suppression away from the peak.}
    \label{fig: bubble3D}
\end{figure*}

Unfortunately, it turns out that for our model neither of these two approaches works in practice.
In fact, the bubble self-energy is very strongly peaked along the coincident diagonal~$r=r'$ and decays exponentially fast away from this line.
In FIG.~\ref{fig: bubble3D} we plot the bubble self-energy in the bounce background as computed with tree-level propagators for~$\alpha=0.7$ and $j=2$.
From the left panel, in linear scale, we see that the bubble is indeed very strongly peaked along the coincident line~$r_x=r_y$. 
On the right panel we see the same plot in logarithmic scale, from which we see the exponential suppression away from the coincident diagonal.
This results in a function of two variables that is really hard to either interpolate or fit with basis functions. 
Thus, attempts at both of these approaches have not led to a result.

\begin{figure*}
    \centering
    \begin{subfigure}[b]{0.45\textwidth}
        \centering
        \includegraphics[width=\textwidth]{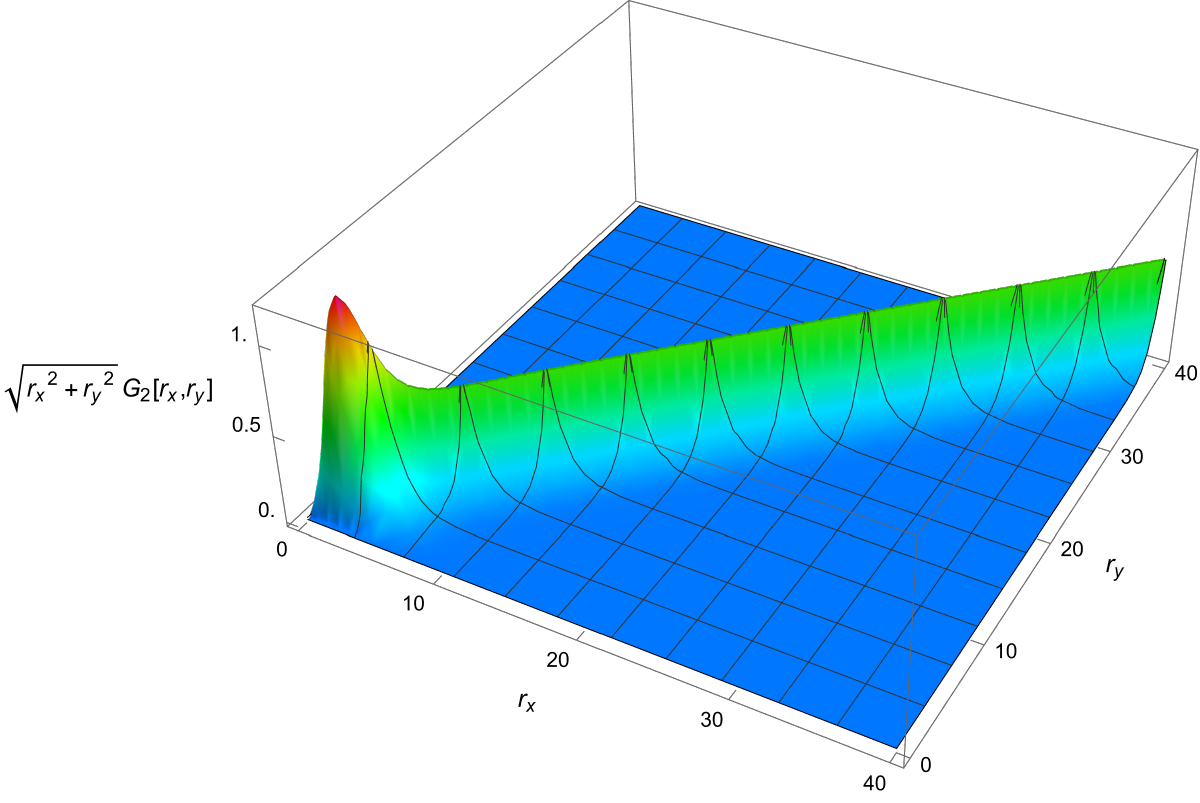}
        \caption{}
    \end{subfigure}
    \hfill
    \begin{subfigure}[b]{0.45\textwidth}
        \centering
        \includegraphics[width=\textwidth]{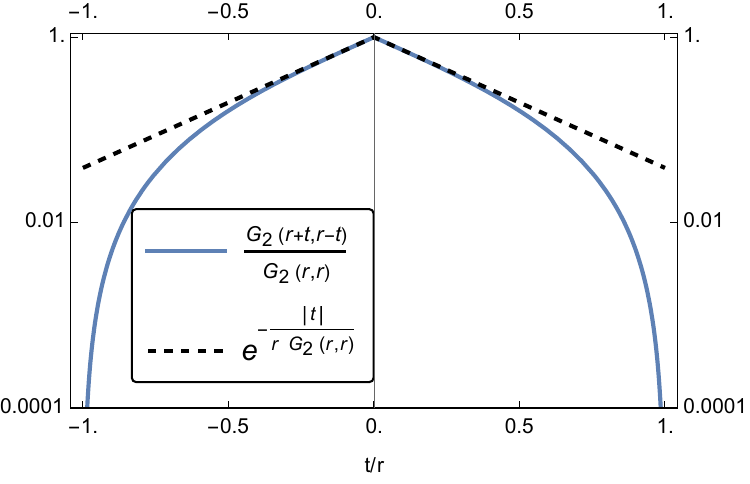}
        \caption{}
    \end{subfigure}
    \caption{{(a)~\bf Left}: tree-level Green's function in the bounce background for~$\alpha=0.7$ and~$j=2$. (b)~{\bf Right:} comparison between the numerical Green's function and the approximated ansatz near the coincident diagonal of Eq.~\eqref{eqn: exp ansatz green}. Here~$r=1$ (b).}
    \label{fig: green3D}
\end{figure*}

However, as it often turns out, we can turn an apparently discouraging obstacle to our advantage.
First of all, we study the behaviour of the Green's function close to the coincident limit,
\begin{align}
    & G_j(r + t, r - t) = \: G_j(r,r) \notag \\
    & \qquad + t \left[ \partial_{r'} G_j(r',r) - \partial_{r'} G_j(r,r') \right]_{r'=r} + \mathcal{O}(t^2) \notag \\
    & \qquad = \: G_j(r,r) - \frac{t}{r} + \mathcal{O}(t^2) \,,
\end{align}
where in the last step we have used that the Green's function is symmetric under the exchange of its arguments, and that the discontinuity in its derivative is fixed by the defining differential equation.
This expansion is consistent with the ansatz
\begin{equation}\label{eqn: exp ansatz green}
    G_j(r+t,r-t) = \: G_j(r,r) e^{-\frac{|t|}{r G_j(r,r)}} \left(1 + \mathcal{O}(t)\right)\,,
\end{equation}
which agrees to great precision with the numerics, as can be seen in the right panel of FIG.~\ref{fig: green3D}.

The convolution in Eq.~\eqref{eqn: green eq 2d} contains integrals of the form
\begin{subequations}
\begin{align}
    &\int_0^\infty \d r'' \, r'' G_\ell (r,r'') G_k (r,r'') G_j (r'',r') \,, \\
    &\int_0^\infty \d r'' \, r'' \phi(r'') G_\ell (r,r'') G_k (r,r'') G_j (r'',r')\,.
\end{align}    
\end{subequations}
In the following, we focus on the first of these two integrals; the same argument as we present applies to the second one.
By Eq.~\eqref{eqn: exp ansatz green}, the integrand is strongly peaked at~$r=r''$, where two of the three Green's functions in the integral have their maximum, and it is exponentially suppressed away from this point.
This suggests that we can approximate such integrals to the leading term in the saddle point approximation.
We therefore expand the integrand to the leading order around the peak
\begin{align}
    & r'' G_\ell (r,r'') G_k (r,r'') G_j (r'',r') \notag \\
    & \approx \: r G_\ell (r,r) G_k(r,r) G_j(r,r') e^{-\frac{|r-r''|}{2r} \left(\frac{1}{G_\ell(r,r)} + \frac{1}{G_k(r,r)} \right)}\,.
\end{align}
The remaining integral can be computed analytically,
\begin{align}
    I_{\ell k}(r) = \: & \int_{r_0}^{r_1} \d r'' \, e^{-\frac{|r-r''|}{2r}f_{\ell k}(r)} \notag \\
    = \: & \frac{2r}{f_{\ell k}(r)} \left( 2 - e^{-\frac{1}{2} \left(\frac{r_1}{r} - 1 \right) f_{\ell k}(r)} - e^{-\frac{1}{2} \left(1 - \frac{r_0}{r} \right) f_{\ell k}(r)} \right)\,,
\end{align}
where
\begin{equation}
    f_{\ell k}(r) = \frac{1}{G_\ell(r,r)} + \frac{1}{G_k(r,r)} \,.
\end{equation}
We have computed the integral with finite boundaries, as they occur in the numerical procedure.
The convolution integral can now be rewritten as
\begin{align}
    \int_0^\infty \d r'' \, r'' G_\ell (r,r'') G_k (r,r'') G_j (r'',r') \approx \: F_{\ell k} (r) G_j(r,r')\,,
\end{align}
where we introduced
\begin{align}
    F_{\ell k} (r) = \: & r G_\ell(r,r) G_k(r,r) I_{\ell k}(r)\,.
\end{align}
We thus see that the bubble contribution $\Sigma_j$ approximately localises in the convolution integral.
It can be checked that this approximation works to the per cent level everywhere but for the boundary region where either~$r$ or~$r'$ is close to~$r_0$. In this region, however, the bubble vanishes exponentially.
Finally, the equation for the Green's function can be rewritten
\begin{align}
    &\hspace{-1em}\Bigg( - \frac{1}{r} \frac{\d}{\d r} r \frac{\d}{\d r} + \frac{j^2}{r^2} + W''(\phi(r)) + \beta \Pi_\phi(r) + \beta \Sigma^{\mathrm{loc}}_{\phi,j}(r) \Bigg) G_j(r,r') \notag \\
    & \qquad\qquad\qquad= \: \frac{1}{r}\delta(r-r') - \delta_{j,1} \phi_{\mathrm{tr}}(r) \,\phi_{\mathrm{tr}} (r') \,,\label{eqn: 2d localised green eom}
\end{align}
where we denoted by~$\Sigma^{\rm loc}_j$ the localised bubble self-energy. It should not be confused with the purely local part of the bubble self-energy as defined in Eq.~\eqref{eqn: bubble non-local plus local} that we have found to be finite in four dimensions. No such contribution is present in two dimensions.
The localised bubble self-energy reads
\begin{align}
    \Sigma^{\mathrm{loc}}_j (r) = \: & \frac{9}{2} \left[ 1 - \alpha \phi(r_x) + \alpha^2 \phi(r_x)^2 \right] \, \mathbf{S}^{\mathrm{loc}}_j (r) \,, \label{eqn: 2d localised bubble}
\end{align}
and
\begin{align}
    2\pi \mathbf{S}^{\mathrm{loc}}_j (r)
    = \: & \left(1- \frac{1}{2} \delta_{j,0} \right) \sum_{\ell = \left\lceil \frac{j+1}{2} \right\rceil}^\infty F_{\ell,|j-\ell|} (r) \notag \\
    & + \frac{1}{2} F_{00}(r) + \frac{1+(-1)^j}{4} F_{\frac{j}{2}\frac{j}{2}}(r) \,.
\end{align}
Note that, as highlighted in Eq.~\eqref{eqn: 2d localised green eom}, we must subtract the translational mode at each step in the iteration procedure, despite the fact that beyond tree-level it is not a zero mode any more. 
In fact, as explained in Section~\ref{sec: decay rate review}, the translational mode is taken out at the level of the saddle point approximation of the path integral, and thus never appears as a fluctuation around the bounce.
More technical details on how to subtract the translational mode even when it is not a zero mode are given in Appendix~\ref{app: subtracting the zero-mode}.
With all of these provisions, we are ready to run our self-consistent procedure until convergence of the result.

\subsection{Results}
The numerical computations are done using \textit{Mathematica} v.14. The shooting algorithm to find the bounce is custom-made, and so is the code to obtain the (subtracted) Green's function for each~$j$.
We find the eigenvalues and eigenfunctions at each step of the iteration procedure using the \textit{Mathematica} built-in function \texttt{NDEigenSystem}.
We plot results for the following choice of parameters
\begin{equation}
    \begin{cases}
        \alpha = 0.7 \\
        \beta = 0.02 
    \end{cases} 
    \equiv
    \quad
    \begin{cases}
        g = 0.424\, m^2 \\
        \lambda = 0.042 \, m^2
    \end{cases}
    \,.
\end{equation}

\begin{figure}
    \centering
    \includegraphics[width=\linewidth]{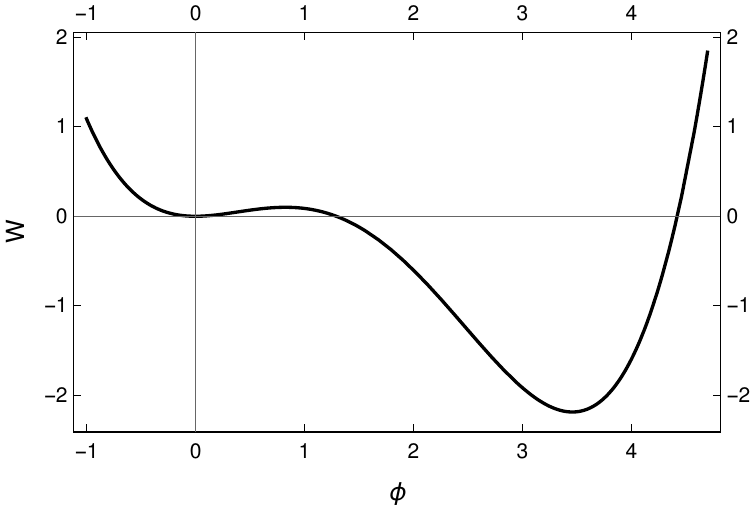}
    \caption{Classical potential~$W(\phi)$ as defined in Eq.~\eqref{eqn: rescaled potential} with~$\alpha=0.7$.}
    \label{fig: potential}
\end{figure}

\begin{figure*}
    \centering
    \begin{subfigure}[b]{0.45\textwidth}
        \centering
        \includegraphics[width=\textwidth]{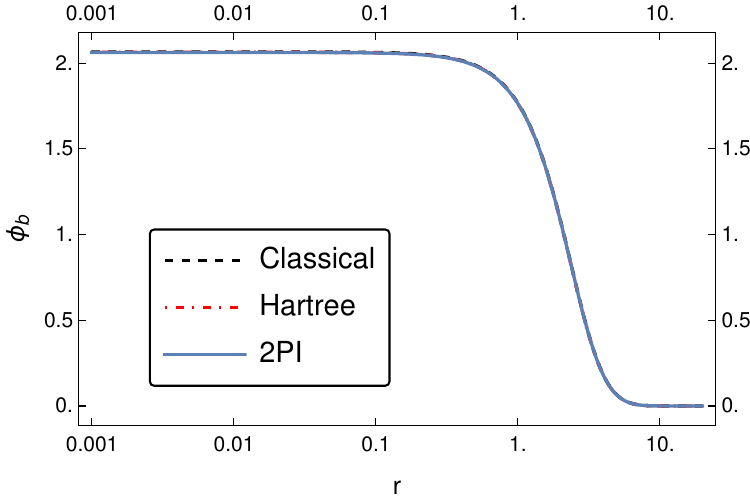}
        \caption{}
    \end{subfigure}
    \hfill
    \begin{subfigure}[b]{0.45\textwidth}
        \centering
        \includegraphics[width=\textwidth]{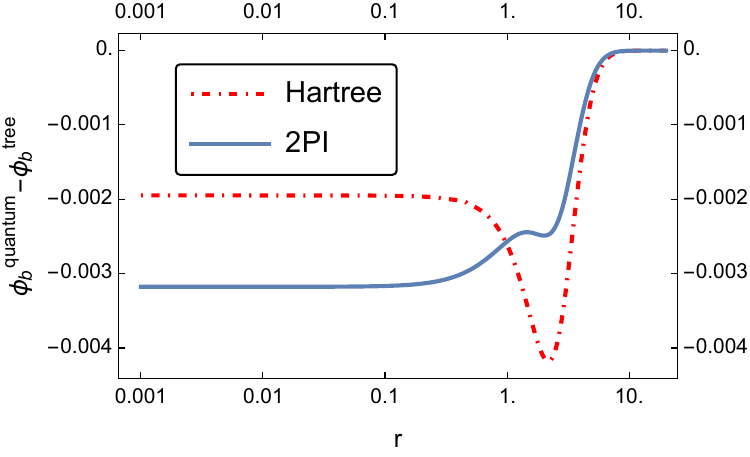}
        \caption{}
    \end{subfigure}
    \caption{(a)~\textbf{Left:} Euclidean bounce at tree-level (dashed), in the Hartree approximation (dotted dashed) and at one-loop in the 2PI formalism (solid). (b)~\textbf{Right:} Difference between the self-consistent quantum bounce and the classical bounce with the quantum solution obtained in the Hartree approximation (dotted dashed) and in full 2PI (solid).}
    \label{fig: bounce}
\end{figure*}

In FIG.~\ref{fig: potential}, we show the potential~$W$ as defined in Eq.~\eqref{eqn: rescaled potential} for the given choice of parameters. In the left panel of FIG.~\ref{fig: bounce}, we plot the classical bounce together with the result of the self-consistent procedure both in the Hartree approximation and in full~2PI.
The corrections to the bounce are minimal, and the action computed on the classical and the quantum bounce agree to the~0.01\%.
The difference between the quantum bounce and the classical bounce, as found in the Hartree approximation and in the full~2PI, is shown in the right panel of FIG.~\ref{fig: bounce}. Though the corrections to the bounce are minimal, we already see a discrepancy between the two approaches.

\begin{figure*}
    \centering
    \begin{subfigure}[b]{0.45\textwidth}
        \centering
        \includegraphics[width=\textwidth]{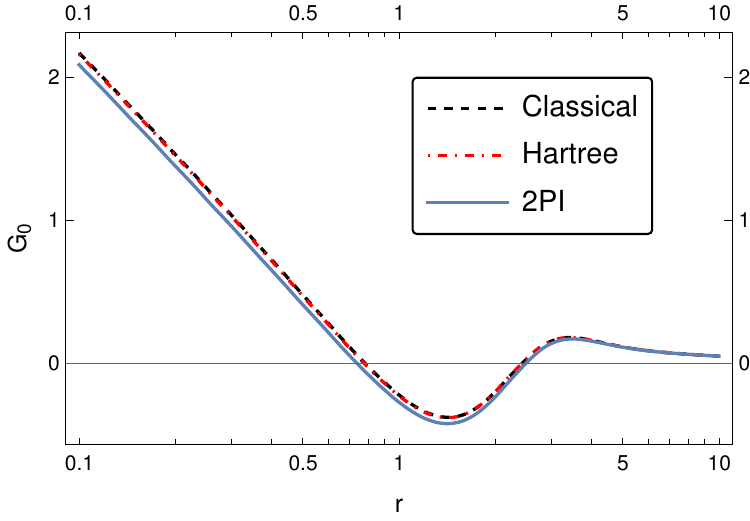}
        \caption{$j=0$}
    \end{subfigure}
    \hfill
    \begin{subfigure}[b]{0.45\textwidth}
        \centering
        \includegraphics[width=\textwidth]{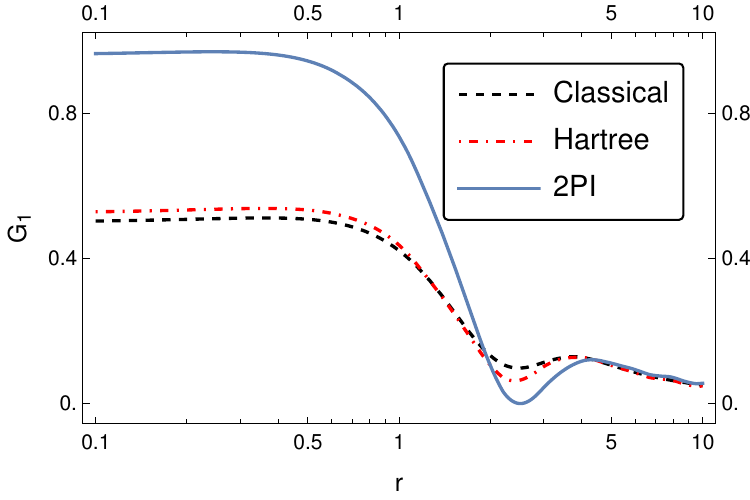}
        \caption{$j=1$}
    \end{subfigure}
    \\[2ex]
    \begin{subfigure}[b]{0.45\textwidth}
        \centering
        \includegraphics[width=\textwidth]{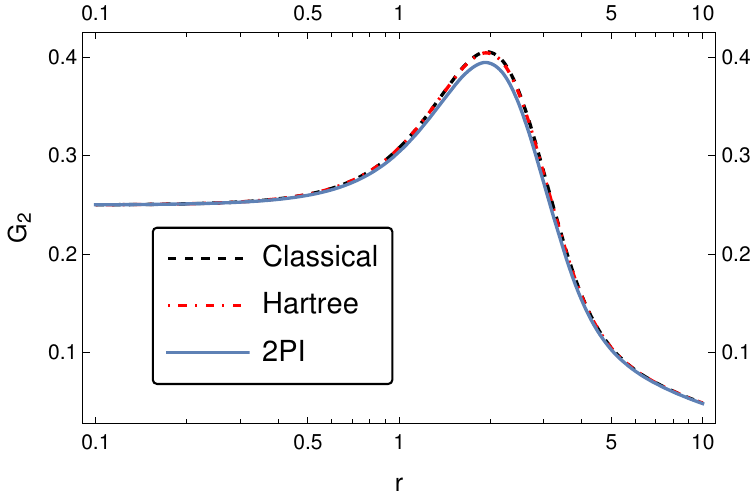}
        \caption{$j=2$}
    \end{subfigure}
    \hfill
    \begin{subfigure}[b]{0.45\textwidth}
        \centering
        \includegraphics[width=\textwidth]{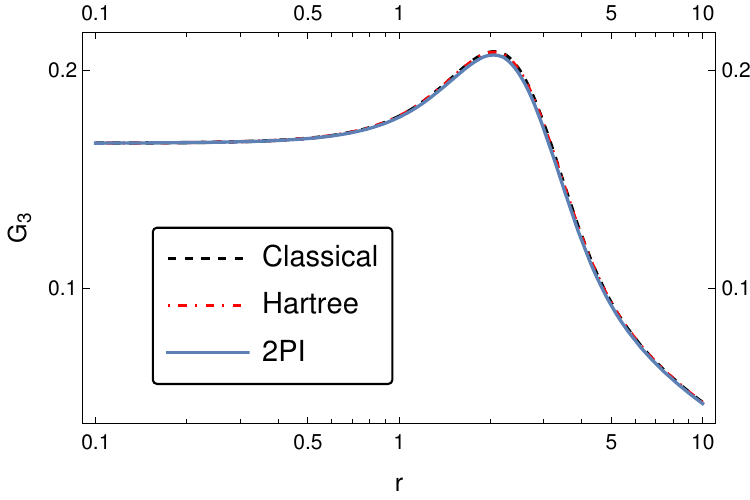}
        \caption{$j=3$}
    \end{subfigure}
    \caption{Coincident Green's function~$G_j(r,r)$ in the bounce background at tree-level (dashed) and as the result of the self-consistent procedure in the Hartree approximation (dotted dashed) and in full 2PI (solid) for~$\alpha=0.7$ and~$\beta=0.02$. The tree-level and Hartree results are on top of each other for all modes except~$j=1$.}
    \label{fig: Green}
\end{figure*}

In FIG.~\ref{fig: Green}, we plot the Green's function in the bounce background. We compare the tree-level with the propagators that include the summation of self-energies evaluated in the Hartree approximation and in full 2PI.
First of all, we observe how the Hartree propagators are basically coincident with the tree-level propagators for all modes but~$j=1$. This is not the case for the 2PI propagators, demonstrating \textit{a posteriori} that keeping the bubble self-energy is crucial.
Looking at the 2PI propagators, the effect of quantum corrections on the Green's function is visible particularly so for the low~$j$ modes.
The effect is particularly pronounced for~$j=1$. 
In this sector resides the translational mode, which we always subtract from the Green's function and that, through the self-consistent procedure, acquires a small positive eigenvalue. 
This is an expression of the fact that Goldstone's theorem only applies approximately once we truncate the effective action to a given loop order, as commented in Section~\ref{sec: decay rate review}.
It will therefore be interesting to reconsider in particular~$j=1$ once we work with the symmetry-improved effective action to ensure that the eigenvalue of translations remains zero throughout the procedure.

Perturbative corrections in the Green's function immediately imply corresponding corrections to the fluctuation determinant and, therefore, to the decay rate. 
This is exemplified by the correction received by the negative eigenvalue
\begin{equation}
    \lambda_{\mathrm{neg}}^{(\mathrm{tree})} = \: -0.58 \,, \qquad \lambda_{\mathrm{neg}}^{(2\mathrm{PI})} = \: -0.54 \,,
\end{equation}
namely, the negative eigenvalue changes by~10\% when including the quantum corrections, compatible with the chosen value for the coupling~${\beta=0.02}$.
In comparison, the Hartree negative eigenvalue changes by about~$0.1$\% compared to the tree-level result.
Though we do not evaluate the decay rate here, this suggests that the decay rate as computed in the Hartree approximation would miss a significant portion of the quantum effects.

\begin{figure*}
    \centering
    \begin{subfigure}[b]{0.45\textwidth}
        \centering
        \includegraphics[width=\textwidth]{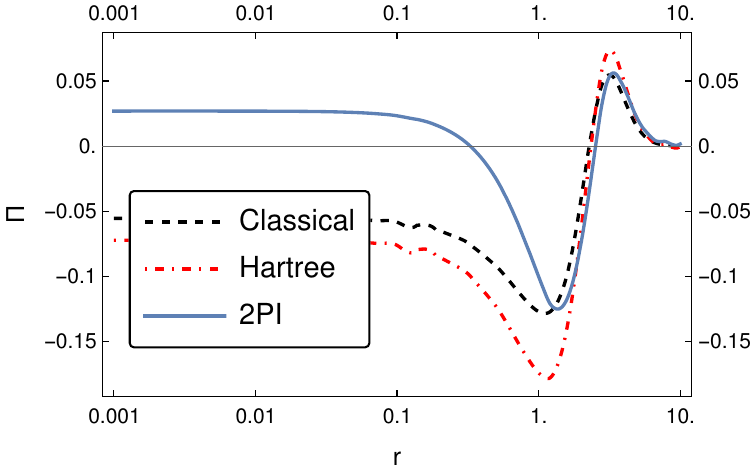}
        \caption{}
    \end{subfigure}
    \hfill
    \begin{subfigure}[b]{0.45\textwidth}
        \centering
        \includegraphics[width=\textwidth]{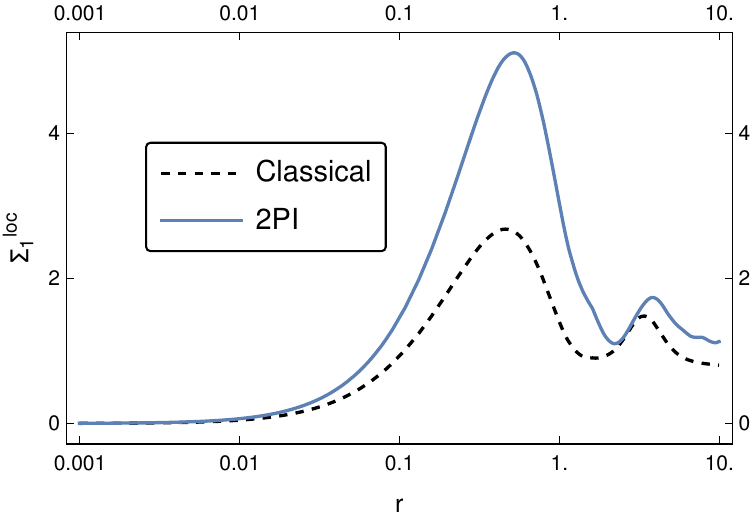}
        \caption{}
    \end{subfigure}
    \caption{(a)~\textbf{Left:} tadpole self-energy as defined in Eq.~\eqref{eqn: tadpole 2d} evaluated with tree-level propagators (dashed) and with summed propagators obtained in the Hartree approximation (dotted dashed) and in full 2PI (solid). (b)~\textbf{Right:} localised bubble self-energy as defined in Eq.~\eqref{eqn: 2d localised bubble} evaluated with tree-level propagators (dashed) and with summed 2PI propagators (solid). By definition, the bubble in the Hartree approximation is always zero.}
    \label{fig: tadpole and bubble}
\end{figure*}

Finally, in FIG.~\ref{fig: tadpole and bubble}, we plot the tadpole self-energy (left panel) and the localised bubble one (right panel), as defined in Eqs.~\eqref{eqn: tadpole 2d} and~\eqref{eqn: 2d localised bubble}, respectively.
Already at tree-level, we see that the two are of the same order and that throwing away the bubble contribution, as done in the Hartree approximation, is generally not justified.
It is interesting to observe how both contributions to the self-energy are strongly modified when using the propagators with the summed-up quantum corrections, suggesting that a loop correction using tree-level propagators hardly leads to an improvement in accuracy, as the extra 2PI or Hartree corrections are of the same order as those based on the tree-level propagators.

\section{Conclusions}\label{sec: conclusions}
Quantum corrections to false vacuum decay may play an essential role, especially in theories with classical symmetries such as massless~$\lambda\phi^4$ with~$\lambda<0$, relevant for the question of the Standard Model metastability.
To perform a qualitative and quantitative analysis, we have re-formulated the problem of finding the decay rate purely in terms of the effective action.
The 2PI effective action formalism is then a natural choice to study systematically the backreaction of fluctuations onto the background and to sum consistently the self-energy insertions in the propagator. 

We have derived equations for the background and for the propagator of a real scalar field in an arbitrary number of dimensions~$d$ with cubic and quartic interaction by assuming~$O(d)$ invariance. 
For the first time, we have computed the non-local contribution to the one-loop self-energy, namely the bubble, in a radially symmetric background. 
We extracted the UV pole of the bubble diagram, and in doing so, we have developed a procedure for how to isolate the local divergences in the position space representation.
The analytic results show that there is, in general, no difference in importance between the tadpole, i.e. the local part of the self-energy, and the bubble, therefore implying that the Hartree approximation is generally not justified.
This we have checked with a parametric example in~$d=2$, for which all qualitative features are present, and convergence of the self-consistent procedure works best.

After outlining the iterative procedure necessary to solve for the background and for the propagator self-consistently, we have observed how we can approximately localise the bubble convolution integral, thus reducing our integro-differential equation to an ordinary non-linear differential equation.

For the given parameters, our results show that, while the background is only very slightly affected by quantum corrections, the same does not apply to the propagator.
The Green's functions receive perturbative corrections when the quantum corrections are summed up within the self-energy, thus implying an equivalent change in the eigenvalues of the fluctuation operator.
This is exemplified in the value of the negative eigenvalue, which, in our parametric example, changes by about~10\% after accounting for quantum corrections.
This has consequences on the value of the fluctuation determinant and, therefore, of the decay rate, whose value will receive corrections once quantum effects are accounted for following the procedure presented in this work.
In particular, we expect the effect to be particularly strong in theories where the one-loop contribution becomes dominant over the classical one, such as in the classically scale invariant model or in theories where the true vacuum is radiatively generated~\cite{Garbrecht:2015yza}.

Having obtained a semi-analytic expression for the renormalised self-energy in dimensional regularisation, with an explicit dependence on the renormalisation scale, we aim to analyse the renormalisation scale dependence of the fluctuation determinant in upcoming work.
In particular, applying our methods to~$d=4$ makes it possible to study the effect of renormalisation on the classically scale invariant Fubini--Lipatov instanton, based on the explicit expression for the breaking of this symmetry in the self-energy.
This will allow us to make a quantitative comparison with the results from Ref.~\cite{Andreassen:2017rzq}.

Furthermore, it is particularly interesting to consider the effect of quantum corrections to the decay rate (or nucleation rate) on the production of gravitational waves during a first order phase transition in the early Universe.
In this work, we have only accounted for quantum corrections at zero temperature, however the same formalism can be extended at finite temperature.
Computing the decay rate beyond the quadratic approximation would reduce the uncertainty on the predicted power spectrum of the produced gravitational waves, thus affecting their projected detectability in future experiments~\cite{Caprini_2020}.

\appendix

\section{The WKB Green's function}\label{app: WKB green}
To find the WKB expansion of the Green's function solution of Eq.~\eqref{eqn: 1 loop 2pi green eom} having dropped the convolution integral, we find it useful to perform a change of variable. We define the variables $s = \log m r$ and $s' = \log m r'$, where $m^2 = U''(\varphi_{\rm FV})$. This follows along the lines of the WKB expansion for the Gel'fand Yaglom functional as carried out in Ref.~\cite{Ekstedt_2023}. The differential equation then becomes
\begin{align}\label{eqn: diff eq exp for green F}
    \left[ - \frac{\d^2}{\d s^2} + E_j^2(s) \right] F_j(s,s') = \: \delta (s-s') \,,
\end{align}
where we have renamed~${G_j (m^{-1}e^s, m^{-1}e^{s'}) = m^2 F_j (s,s')}$, we have defined 
\begin{equation}
    E_j^2(s) = \: j^2 + e^{2s} \frac{m_\varphi^2(e^s)}{m^2} \,,
\end{equation}
with~${m_\varphi^2(r) = U''(\varphi(r)) + \Pi_\varphi(r)}+\varsigma_\varphi(r)$ and everywhere we made the replacement ~${j+\kappa \longrightarrow j}$.
Equation~\eqref{eqn: diff eq exp for green F} lends itself to WKB approximation for large~$E_j^2$, equivalent to the large~$j$ expansion. We make the following ansatz for the Green's function,
\begin{equation}
    F_j (s,s') = \: \frac{1}{\mathcal{W}[f_j^>(s'),f_j^<(s')]} f_j^>(s_>) f_j^< (s_<) \,,
\end{equation}
where ${s_> = \max (s,s')}$ and analogously ${s_< = \min (s,s')}$ and~${\mathcal{W}[f,g]=fg'-f'g}$ is the Wronskian of~$f$ and~$g$. The two functions~$f^{\lessgtr}$ are the left and right-regular solutions, respectively, defined by the equations
\begin{equation}
\begin{cases}
    \left[ - \frac{\d^2}{\d s^2} + E_j^2(s) \right] f^{\lessgtr}(s) = 0 \,, \\[1.5ex]
    \lim_{s \to \mp \infty} f^{\lessgtr}(s)  = 0 \,.
\end{cases}
\end{equation}
We are ready to make the WKB ansatz. First, we introduce a counting parameter $\epsilon$ that we will later set to one. We want to find the WKB solution to 
\begin{equation}
    \left[ - \epsilon^2 \frac{\d^2}{\d s^2} + E_j^2(s) \right] f_j^{\lessgtr}(s) = 0 \,,
\end{equation}
via the WKB ansatz
\begin{equation}
    f_j^{\lessgtr}(s) = \: e^{\frac{1}{\epsilon} \sum_{n=0}^\infty \epsilon^n h_n^{(j)}(s) } = \: e^{h_j^{\lessgtr}(s)}\,.
\end{equation}
Solving the equation up to order $\epsilon^2$ we find the WKB-expanded Green's function
\begin{equation}
    F_j(s,s') = \: \frac{1}{2\omega_j(s')} \sqrt{\frac{E_j(s')}{E_j(s)}} e^{-\int_{s_<}^{s_>} \d \tilde s \, \omega_j(\tilde s)} \,,
\end{equation}
where up to order $\epsilon^2$ the function $\omega$ reads
\begin{equation}
    \omega_j(s) = E_j(s) + \frac{-3E_j'(s)^2 + 2E_j(s)E_j''(s)}{8E_j(s)^3} \,.
\end{equation}
Here, the fraction is suppressed for large $E_j$. We therefore proceed now to make a large $j$ expansion
\begin{widetext}
\begin{subequations}    
\begin{align}
    E_j(s) = \: & j + \frac{e^{2s} m_\varphi(e^s)^2}{2j m^2} - \frac{e^{4s} m_\varphi(e^s)^4}{8j^3 m^4} + \mathcal{O}(j^{-5}) \,, \\
    \frac{1}{2\omega_j(s)} = \: & \frac{1}{2j} - \frac{e^{2s}m_\varphi(e^s)^2}{4j^3 m^2} + \mathcal{O}(j^{-5}) \,, \\
    \int_{s_<}^{s_>} \hspace{-0.5em} \d\tilde s \, \omega_j(\tilde s) = \: & j (s_> - s_<) + \frac{1}{2j} \int_{s_<}^{s_>} \hspace{-0.5em} \d \tilde s \, e^{2\tilde s} \frac{m_\varphi(e^{\tilde s})^2}{m^2} + \mathcal{O}(j^{-3}) \,.
\end{align}
\end{subequations}
Thus, we can write the Green's function expanded for large $j$ as
\begin{align}
    G_{j-\kappa}(r,r') = \: & \frac{1}{2j} \left(1 - \frac{3 r^2 m_\varphi^2(r) - r'^2 m_\varphi(r')^2}{4j^2} + \mathcal{O}(j^{-4}) \right) \left( \frac{r_<}{r_>} \right)^{j} \left(1 - \frac{A (r_>,r_<)}{2j} + \frac{A (r_>,r_<)^2}{8j^2} + \mathcal{O}(j^{-3}) \right) \notag \\
    = \: & \frac{1}{2j} \left( \frac{r_<}{r_>} \right)^{j} \left(1 - \frac{A (r_>,r_<)}{2j} + \frac{A (r_>,r_<)^2}{8j^2} - \frac{3 r^2 m_\varphi^2(r) - r'^2 m_\varphi(r')^2}{4j^2}  + \mathcal{O}(j^{-3}) \right)\,,
\end{align}
having defined
\begin{equation}
    A (r_>,r_<) = \: \int_{r_<}^{r_>} \d \tilde r \, \tilde r \, \frac{m_\varphi^2 (\tilde r) }{m^2} \,. 
\end{equation}

\section{Linearising the square of the Green's function}\label{app: linearising square green}
Starting from Eq.~\eqref{eqn: definition linear green squared} and using the linearisation formula for the Gegenbauer polynomials~\eqref{eqn: Gegenbauer linearisation formula}, we want to identify the term~$\mathbf{S}_j$.
First, we give the linearisation coefficients appearing in Eq.~\eqref{eqn: Gegenbauer linearisation formula}
    \begin{equation}
        \gamma_{jj'p}^{(\kappa)} = \: \frac{j+j'-2p+\kappa}{j+j'-p+\kappa} \frac{\Gamma(j+j'-2p+1)}{\Gamma(p+1)\Gamma(j-p+1)\Gamma(j'-p+1)} \frac{(\kappa)_p (\kappa)_{j-p} (\kappa)_{j'-p} (2\kappa)_{j+j' - p}}{(\kappa)_{j+j'-p} (2\kappa)_{j+j'-2p}}\,. \label{eqn: linearisation coefficients}
    \end{equation}
\end{widetext}
We note that this is identically one if~${\kappa=1}$, namely~${d=4}$.
We start by using the linearisation formula~\eqref{eqn: Gegenbauer linearisation formula} to write
\begin{equation}\label{eqn: linearisation before redefinition}
    G(x,y)^2 = \: a_\kappa^2 \sum_{j,j' = 0}^\infty \sum_{p=0}^{\min(j,j')} \alpha^{(\kappa)}_{j,j',p} C^\kappa_{j+j'-2p}\,, 
\end{equation}
having defined the parameter
\begin{equation}
    a_\kappa = \: \frac{\Gamma(\kappa)}{2\pi^{\kappa+1} (r_x r_y)^\kappa}\,,
\end{equation}
and where~$\alpha_{j,j',p}^{(\kappa)}$ has been defined in Eq.~\eqref{eqn: definition alphas}.
Now, we split the sum in Eq.~\eqref{eqn: linearisation before redefinition} into three contributions,
\begin{align}
    G(x,y)^2 = \: & \underbrace{a_\kappa^2 \sum_{j>j'} \sum_{p=0}^{j'} \alpha^{(\kappa)}_{j,j',p} C^\kappa_{j+j'-2p}}_{(a)} + \underbrace{(j\leftrightarrow j')}_{(a')} \notag \\
    & + \underbrace{a_\kappa^2 \sum_{j=0}^\infty \sum_{p=0}^{j} \alpha^{(\kappa)}_{j,j,p} C^\kappa_{2j-2p}}_{(b)} \,. \label{eqn: green squared split in three sums}
\end{align}
Because~$j$ and~$j'$ are just mute indices, terms~$(a)$ and~$(a')$ are exactly the same.

\paragraph*{$(a)$-term} To rewrite the sum~\eqref{eqn: green squared split in three sums} in the desired form, we introduce the index~${\ell=j+j'-2p}$. In the sum, we change from index~$p$ to index~$\ell$.
To obtain the correct summation boundaries, we find it useful to rewrite the sums without any boundaries and introduce~$\theta$-functions to enforce them. Omitting the argument of the sum, which is not important for determining the boundaries, we have
\begin{align}
    \sum_{p=0}^{j'} & = \: \sum_{p \in \mathbb{Z}} \sum_{\ell\in \mathbb{Z}} \theta_p \theta_{j'-p} \delta_{\ell,j+j'-2p} \notag \\
    & = \: \sum_{\ell\in \mathbb{Z}} \theta_{\frac{j+j'-\ell}{2}} \theta_{\frac{j'-j+\ell}{2}} \frac{1+(-1)^{j+j'-\ell}}{2} \notag \\
    & = \: \sum_{\ell=j-j'}^{j+j'} \frac{1+(-1)^{j+j'-\ell}}{2}\,,
\end{align}
where we are enforcing explicitly that~${2p = j+j'-\ell}$ must be even.
Next, we introduce the indices~${j_{\pm} = j \pm j'}$, and use them to replace~$j$ and~$j'$
\begin{align}
    \sum_{j=1}^\infty\sum_{j'=0}^{j-1} = \: & \sum_{j,j'\in\mathbb{Z}} \sum_{j_{\pm}\in \mathbb{Z}} \theta_{j'} \theta_{j-1-j'}\theta_{j-1}\delta_{j_-,j-j'} \delta_{j_+,j+j'} \notag \\
    = \: & \sum_{j_\pm \in \mathbb{Z}} \theta_{j_+ - j_-} \theta_{j_- - 1} \theta_{j_+ + j_- -2} \frac{1+(-1)^{j_+ + j_-}}{2} \notag \\
    = \: & \sum_{j_+=1}^\infty \sum_{j_- = 1}^{j_+} \frac{1+(-1)^{j_+ + j_-}}{2} \,,
\end{align}
where once again, we must impose that~${j_+ + j_-}$ is even.
We can then rewrite term~$(a)$ in Eq.~\eqref{eqn: green squared split in three sums} as
\begin{align}
    (a) = \: & a_\kappa^2 \sum_{j_+\geq j_- =1}^\infty \sum_{\ell=j_-}^{j_+} \frac{1+(-1)^{j_+ +\ell}}{2} \frac{1+(-1)^{j_+ + j_-}}{2} \times \notag \\
    & \times \alpha_{\frac{j_++j_-}{2},\frac{j_+-j_-}{2},\frac{j_+-\ell}{2}} C_\ell^\kappa \,.
\end{align}
Our goal is to pull out a sum over~$\ell$, so we rearrange the sum as follows,
\begin{equation}
    \sum_{j_+\geq j_- =1}^\infty \sum_{\ell=j_-}^{j_+} = \: \sum_{j_+\geq \ell \geq j_- \geq1} = \: \sum_{\ell = 1}^\infty \sum_{j_+ = \ell}^\infty \sum_{j_- =1}^\ell\,,
\end{equation}
so that, after the re-labelling~$\ell\to j$, we can define~$\mathbf{S}_j^{(a)}$ as
\begin{equation}
    (a) = \: a_\kappa \sum_{j=0}^\infty (j+\kappa) \mathbf{S}_j^{(a)}C_j^\kappa\,,
\end{equation}
and identify
\begin{align}
    \mathbf{S}_j^{(a)} = \: & \frac{a_\kappa}{j+\kappa} \sum_{j_+ = j}^\infty \sum_{j_- =1}^j  \frac{1+(-1)^{j_+ +j}}{2} \frac{1+(-1)^{j_+ + j_-}}{2} \times \notag \\
    & \times \alpha_{\frac{j_++j_-}{2},\frac{j_+-j_-}{2},\frac{j_+-j}{2}} \,.
\end{align}
Note that we should have no term for~$j=0$. To avoid introducing additional terms when~$j=0$, we understand the following sum to vanish
\begin{equation}
    \sum_{j_-=1}^0 = \: 0\,.
\end{equation}
Now we are almost there. 
We call~$\ell=\frac{j_++j_-}{2}$, and replace the sum over~$j_+$
\begin{equation}
    \sum_{j_+=j}^\infty = \: \sum_{j_+,\ell\in\mathbb{Z}} \theta_{j_+-j} \delta_{\ell,\frac{j_++j_-}{2}} = \: \sum_{\ell=\frac{j_-+j}{2}}^\infty\,,
\end{equation}
where we used that~$j_++j_-$ is even. This, together with the constraint that~${j_++j}$ must be even, implies that so is~${j_-+j}$ as well.
What we have at this stage is
\begin{align}
    \mathbf{S}_j^{(a)} = \: & \frac{a_\kappa}{j+\kappa} \sum_{j_- =1}^j \sum_{\ell=\frac{j_-+j}{2}}^\infty   \frac{1+(-1)^{j_- +j}}{2} \times \notag \\
    & \times \alpha_{\ell,\ell - j_-,\ell - \frac{j_-+j}{2}} \,.
\end{align}
For our final substitution, we define~${q=\frac{j_-+j}{2}}$ and replace the sum over~$j_-$,
\begin{equation}
    \sum_{j_-=1}^j = \: \sum_{j_-,q\in\mathbb{Z}} \theta_{j_--1} \theta_{j-j_-} \delta_{q,\frac{j_-+j}{2}} = \: \sum_{q=\left\lceil \frac{j+1}{2} \right\rceil}^j \,,
\end{equation}
where the ceiling function arises from requiring that~${2q\geq j+1}$, and that~$q$ is an integer.
After applying this last transformation and substituting
\begin{equation}
    \left\lceil \frac{j+1}{2} \right\rceil = \left\lfloor \frac{j+2}{2} \right\rfloor\,,    
\end{equation}
we obtain the first term in Eq.~\eqref{eqn: linearised green squared}, where the factor of two comes from the contribution of the~$(a')$ term in the sum~\eqref{eqn: green squared split in three sums}.

\paragraph*{$(b)$-term}
We now turn to the~$(b)$-term in Eq.~\eqref{eqn: green squared split in three sums}.
This is much easier to deal with.
We start by introducing the index~${k=2j-2p}$, to get rid of the~$p$ sum,
\begin{equation}
    \sum_{j\geq p =0}^\infty = \sum_{j,p,k\in\mathbb{Z}} \theta_{j-p}\theta_p \delta_{k,2j-2p} = \: \sum_{k = 0}^\infty \sum_{j=\frac{k}{2}}^\infty \frac{1+(-1)^{k}}{2} \,.
\end{equation}
Renaming~$k\to j$ and~$j\to\ell$ we find the last term in Eq.~\eqref{eqn: linearised green squared}.

\section{The bubble self-energy in $d=2$}\label{app: bubble in d=2}
In using the expressions~\eqref{eqn: definition linear green squared} and~\eqref{eqn: linearised green squared}, we need to be careful, as they both appear to be problematic when~$\kappa\to0$.
In order to take the limit smoothly, we first employ the following identity
\begin{equation}\label{eqn: gegenbauer in terms of jacobi}
    C_j^\kappa (\cos\theta) = \: \frac{(2\kappa)_j}{\left(\kappa+\frac{1}{2}\right)_j} P_j^{\left(\kappa-\frac{1}{2},\kappa-\frac{1}{2}\right)} (\cos\theta)\,,
\end{equation}
where~$P_j^{(a,b)}$ are the Jacobi polynomials, which for~$\kappa\to0$ reduce to Chebyshev polynomials of the first kind,
\begin{equation}
    P_j^{\left(-\frac{1}{2},-\frac{1}{2}\right)} (\cos\theta) = \: \frac{\left(\frac{1}{2}\right)_j}{j!} T_j(\cos\theta)\,.
\end{equation}
Equation~\eqref{eqn: gegenbauer in terms of jacobi} reveals that the Gegenbauer polynomials have a simple zero for~$\kappa\to0$, which simplifies with the simple pole of~$\Gamma(\kappa)$ appearing in Eq.~\eqref{eqn: definition linear green squared}.
Then, the latter equation for~$\kappa\to0$ reduces to
\begin{equation}
    G(x,y)^2 = \: \frac{1}{\pi} \sum_{j=0}^\infty T_j(\cos\theta) \mathbf{S}_j(r_x,r_y)\,.
\end{equation}
Note that the same care must be taken when writing the decomposition of the Green's function as in Eq.~\eqref{eqn: definition Green's fnct in angular decomposition}. We obtain
\begin{equation}
    G(x,y) = \: \frac{1}{\pi} \sum_{j=0}^\infty T_j(\cos\theta) G_j(r_x,r_y)\,.
\end{equation}
To get the smooth limit of Eq.~\eqref{eqn: linearised green squared}, we first look at the linearisation formula for the product of Chebyshev polynomials
\begin{align}
    T_j(\cos\theta) T_{j'}(\cos\theta) = \: & \sum_{p=0}^{\min(j,j')} \gamma_{j,j',p} T_{j+j'-2p} (\cos\theta) \notag \\
    = \: & \frac{1}{2} \left( T_{j+j'}(\cos\theta) + T_{|j-j'|} \right)\,, \label{eqn: linearisation formula chebyshev}
\end{align}
which implies that the linearisation coefficients read
\begin{equation}
    \gamma_{j,j',p} = \: \frac{1}{2} \left( \delta_{p,0} + \delta_{p,\min(j,j')}\right) \,.
\end{equation}
Thus, we understand that the limit of the linearisation coefficients in Eq.~\eqref{eqn: linearisation coefficients} should reproduce this, namely
\begin{equation}
    \lim_{\kappa\to0} \gamma^{(\kappa)}_{j,j',p} = \: f(\kappa) \left( \delta_{p,0} + \delta_{p,\min(j,j')}\right) \,,
\end{equation}
where~$f(\kappa)$ is some function of~$\kappa$ that vanishes when~$\kappa$ goes to zero.
We can verify this by expanding the linearisation coefficients~\eqref{eqn: linearisation coefficients} around~$\kappa=0$, multiplying by~$\Gamma(\kappa)$, which is the only other possibly singular term in Eq.~\eqref{eqn: linearised green squared},
\begin{align}
    \Gamma(\kappa) \gamma^{(\kappa)}_{j,j',p} = \: & \frac{(j+j'-2p)^2}{(j-p)(j'-p)(j+j'-p)p} \kappa + \mathcal{O}(\kappa^2) \,, \label{eqn: expanded Gamma times gamma for k to 0}
\end{align}
where we recall that~$p$ goes from zero to~$\min(j,j')$.
When~$\kappa=0$, Eq.~\eqref{eqn: expanded Gamma times gamma for k to 0} can only be non-zero if either~$p=0$, or~$p=\min(j,j')$.
Thus we find
\begin{equation}
    \lim_{\kappa\to0} \Gamma(\kappa) \gamma^{(\kappa)}_{j,j',p} = \: \frac{j+j'-2p}{jj'} \frac{1}{2} \left( \delta_{p,0} + \delta_{p,\min(j,j')}\right)\,.
\end{equation}
Computing the limit of the remainder of the terms in Eq.~\eqref{eqn: linearised green squared} is straightforward, and we finally find
\begin{align}
    2\pi \mathbf{S}_j 
    = \: & \left(1- \frac{1}{2} \delta_{j,0} \right) \sum_{\ell = \left\lceil \frac{j+1}{2} \right\rceil}^\infty G_\ell G_{|j-\ell|} \notag \\
    & + \frac{1}{2} G_0^2 + \frac{1+(-1)^j}{4} G_{\frac{j}{2}}^2 \,.
\end{align}
By using the linearisation formula~\eqref{eqn: linearisation formula chebyshev} instead, we can work out the same object to arrive at the same result.
This provides an independent check of Eq.~\eqref{eqn: linearised green squared}.
From Eq.~\eqref{eqn: WKB green full} we know that~$G_\ell \sim 1/\ell$ for large angular momentum~$\ell$, which immediately implies that the infinite sum in Eq.~\eqref{eqn: linearised green squared 2d} is finite even in the coincident limit.
This should not come as a surprise: we know the bubble to be finite in~$d=2$.

\section{Subtracting the translational mode}\label{app: subtracting the zero-mode}
As explained in Section~\ref{sec: decay rate review}, the translational mode is not a propagating degree of freedom of the theory, and it must therefore be subtracted from the Green's function.
In particular, we construct the Green's function as the inverse of~$G^{-1}$ as appearing in Eq.~\eqref{eqn: 2pi inverse green eq}, where we invert the propagator only on the subspace of the Hilbert space orthogonal to the translational modes. 
In equations,
\begin{equation}
    G^{-1} G = \: \mathbf{1}^\perp = \: \mathbf{1} - \mathbf{P}_{\mathrm{transl}} \,,
\end{equation}
where~$\mathbf{P}_{\mathrm{transl}}$ is the projection operator onto the span of the translational modes~$\phi_i$.
Thus we find Eq.~\eqref{eqn: 2pi eq G full}.

Subtracting an eigenmode from the propagator is, in principle, very simple.
In the spectral decomposition, we have
\begin{equation}
    G_{\rm full}(x,y) = \: \sum_{\lambda} \frac{\phi_\lambda(x) \phi_\lambda^*(y)}{\lambda} \,,
\end{equation}
where the sum runs over all eigenvalues~$\lambda$, and~$\phi_\lambda$ is the eigenfunction with eigenvalue~$\lambda$, namely
\begin{equation}
    \int \d^d y \left(G_{0,\varphi}^{-1} (x,y) +  \mathbf{\Sigma}_\varphi(x,y) \right) \phi_\lambda(y) = \: \lambda \phi_\lambda(x)\,. 
\end{equation}
Then, subtracting the eigenmode~$\phi_{\lambda_0}$ from the full Green's function amounts to a simple operation 
\begin{equation}
    G(x,y) = \: G_{\rm full}(x,y) - \frac{\phi_{\lambda_0}(x) \phi_{\lambda_0}^*(y)}{\lambda_0} \,.
\end{equation}
However, the issue is very clear: what if~$\lambda_0=0$ or very small, as is the case for the translational mode?
Then we must do something more sophisticated as explained in Ref.~\cite{Garbrecht:2018rqx}, and as we will outline in the remainder of this appendix.

First of all, we reduce ourselves to Eq.~\eqref{eqn: 1 loop 2pi green eom}, where the angular decomposition has already been done. We need to only consider the equation for~$j=1$, where the subtraction is necessary.
We look at the equation for~${r< r'}$
\begin{align}
    &\left( - \frac{1}{r} \frac{\d}{\d r} r \frac{\d}{\d r} + \frac{(1+\kappa)^2}{r^2} + U''(\varphi(r)) + \Pi_\varphi(r) \right) G^<_1(r,r') \notag \\
    &\quad + \int_0^\infty \d r'' \, r'' \, \Sigma_{\varphi,1}(r,r'') \, G^<_1(r'',r') = \:  \phi_0(r) \, \phi_0 (r') \,,
\end{align}
where we defined the zero mode~$\phi_0(r)=r^\kappa\phi_{\mathrm{tr}}(r)$.
The boundary conditions on the Green's function are such that it must approach the false-vacuum Green's function for~$r\to0$. 
Then, the left Green's function can be written as
\begin{equation}
    G^<_1(r,r') = \: L_1(r)\phi_0(r') + \phi_0(r)c^<(r')\,,
\end{equation}
where~$L_1(r)$ is a solution to the non-homogeneous equation
\begin{align}
    &\left( - \frac{1}{r} \frac{\d}{\d r} r \frac{\d}{\d r} + \frac{(1+\kappa)^2}{r^2} + U''(\varphi(r)) + \Pi_\varphi(r) \right) L_1(r) \notag \\
    &\quad + \int_0^\infty \d r'' \, r'' \, \Sigma_{\varphi,1}(r,r'') \, L_1(r'') = \:  \phi_0(r)\,,
\end{align}
satisfying the left-regular boundary conditions.
Note that we can always add a term proportional to the zero mode~$\phi_0(r)$ multiplied by some function of~$r'$. 
Analogously for the right Green's function
\begin{equation}
    G^>_1(r,r') = \: R_1(r) \phi_0(r') + \phi_0(r) c^>(r') \,,
\end{equation}
with regular boundary conditions at~$r\to\infty$.
Imposing symmetry of the Green's function under the exchange of its arguments, we can construct the general solution for the subtracted Green's function 
\begin{align}
    G_1^{\perp} (r,r') = \: & \left[ L_1(r_<)\phi_0(r_>) + \phi_0(r_<) R_1(r_>) \right] \notag \\[1.3ex]
    & + a\phi_0(r) \phi_0(r') \,,
\end{align}
where~$a$ is a numerical constant, and~${r_>=\max (r,r')}$ and~${r_< = \min (r,r')}$.
This solution satisfies the required properties of the Green's function, namely continuity at~$r=r'$ and jump in the derivative at~$r=r'$ of magnitude~$1/r'$.
The coefficient~$a$ can be fixed by demanding that the Green's function is orthogonal to the zero mode, namely
\begin{equation}
    \int_0^\infty \d r \, r \, \phi_0(r) G_1^\perp(r,r') = \: 0 \,.
\end{equation}
We find
\begin{align}
    a=&\int_0^{r'} \d r \, r L(r)\phi_0(r)+\int_{r'}^\infty \d r \, r R(r)\phi_0(r) \notag \\
    &+\frac{L(r')}{\phi_0(r')}\int_{r'}^\infty \d r \, r \, \phi_0(r)^2 + \frac{R(r')}{\phi_0(r')}\int_0^{r'}\d r \, r \,\phi_0(r)^2\,.
\end{align}
Although there seems to be an explicit dependence on~$r'$, the coefficient is completely independent of it, as can be checked numerically. Thus, we are free to choose a convenient~$r'$ when computing~$a$ numerically.

\bibliography{apssamp}

\end{document}